\documentclass[apj,iop]{emulateapj}

\usepackage{natbib}
\usepackage{longtable}
\usepackage{hyperref}

\newcommand{\HI}{H{\footnotesize I}} 
\newcommand{\HII}{H{\footnotesize II}}

\newcommand{\NII}{[N~{\footnotesize II}]}

\newcommand{\OI}{[O~{\footnotesize I}]} 
\newcommand{\OIII}{[O~{\footnotesize III}]} 
\newcommand{\SII}{[S~{\footnotesize II}]}

\shorttitle{Wandering BHs in Dwarf Galaxies}
\shortauthors{Reines et al.}

\begin{document}

\title{A New Sample of (Wandering) Massive Black Holes in Dwarf Galaxies \\ from High Resolution Radio Observations}

\author{Amy E. Reines}
\affil{eXtreme Gravity Institute, Department of Physics, Montana State University, Bozeman, MT 59717, USA}
\email{amy.reines@montana.edu}

\author{James J. Condon}
\affil{National Radio Astronomy Observatory, Charlottesville, VA 22903, USA}

\author{Jeremy Darling}
\affil{Center for Astrophysics and Space Astronomy, Department of Astrophysical and Planetary Sciences, University of Colorado, 389 UCB, Boulder, CO 80309-0389, USA}

\and 

\author{Jenny E. Greene}
\affil{Department of Astrophysical Sciences, Princeton University, Princeton, NJ 08544, USA}

\begin{abstract}

We present a sample of nearby dwarf galaxies with radio-selected accreting massive black holes (BHs), the majority of which are non-nuclear.  We observed 111 galaxies using sensitive, high-resolution observations from the Karl G.\ Jansky Very Large Array (VLA) in its most extended A-configuration at X-band ($\sim$8--12 GHz), yielding a typical angular resolution of $\sim$ 0\farcs25 and rms noise of $\sim 15~\mu$Jy.  Our targets were selected by cross matching galaxies with stellar masses $M_\star \lesssim 3 \times 10^9~M_\odot$ and redshifts $z<0.055$ in the NASA-Sloan Atlas with the VLA Faint Images of the Radio Sky at Twenty centimeters (FIRST) Survey.  With our new high-resolution VLA observations, we detect compact radio sources towards 39 galaxies and carefully evaluate possible origins for the radio emission including thermal \HII\ regions, supernova remnants, younger radio supernovae, background interlopers, and AGNs in the target galaxies.  We find that 13 dwarf galaxies almost certainly host active massive BHs despite the fact that only one object was previously identified as having optical signatures of an AGN.  We also identify a candidate dual radio AGN in a more massive galaxy system.  The majority of the radio-detected BHs are offset from the center of the host galaxies with some systems showing signs of interactions/mergers.  Our results indicate that massive BHs need not always live in the nuclei of dwarf galaxies, confirming predictions from simulations. Moreover, searches attempting to constrain BH seed formation using observations of dwarf galaxies need to account for such a population of ``wandering" BHs. 

\end{abstract}


\section{Introduction}\label{sec:intro}

Massive black holes (BHs)\footnote{Here the term ``massive BH" refers to a non-stellar BH.  Massive BHs have so far been detected in the range of $M_{\rm BH} \sim 10^4 - 10^{10} M_\odot$.} typically reside in the centers of massive galaxies, yet the secrets of BH
formation have been erased in these systems by merger-driven growth over cosmic history \citep[e.g.,][]{volonteri2010,natarajan2014}.
Moreover, the first ``seed" BHs at high redshift \citep[e.g.,][]{latifferrara2016}, that can ultimately grow to millions or billions of solar masses, are too small and faint to detect with existing telescopes (e.g., \citealt{volonterireines2016,vitoetal2018,schleicher2018}).  
A promising alternative approach to learn about BH origins is to find and study the least-massive BHs in nearby dwarf galaxies, which places the most definitive constraints on the masses of BH seeds (see \citealt{reinescomastri2016} for a review).

\begin{figure*}[!t]
\begin{center}$
\begin{array}{cc}
\hspace{-.4cm}
\includegraphics[width=3.6in]{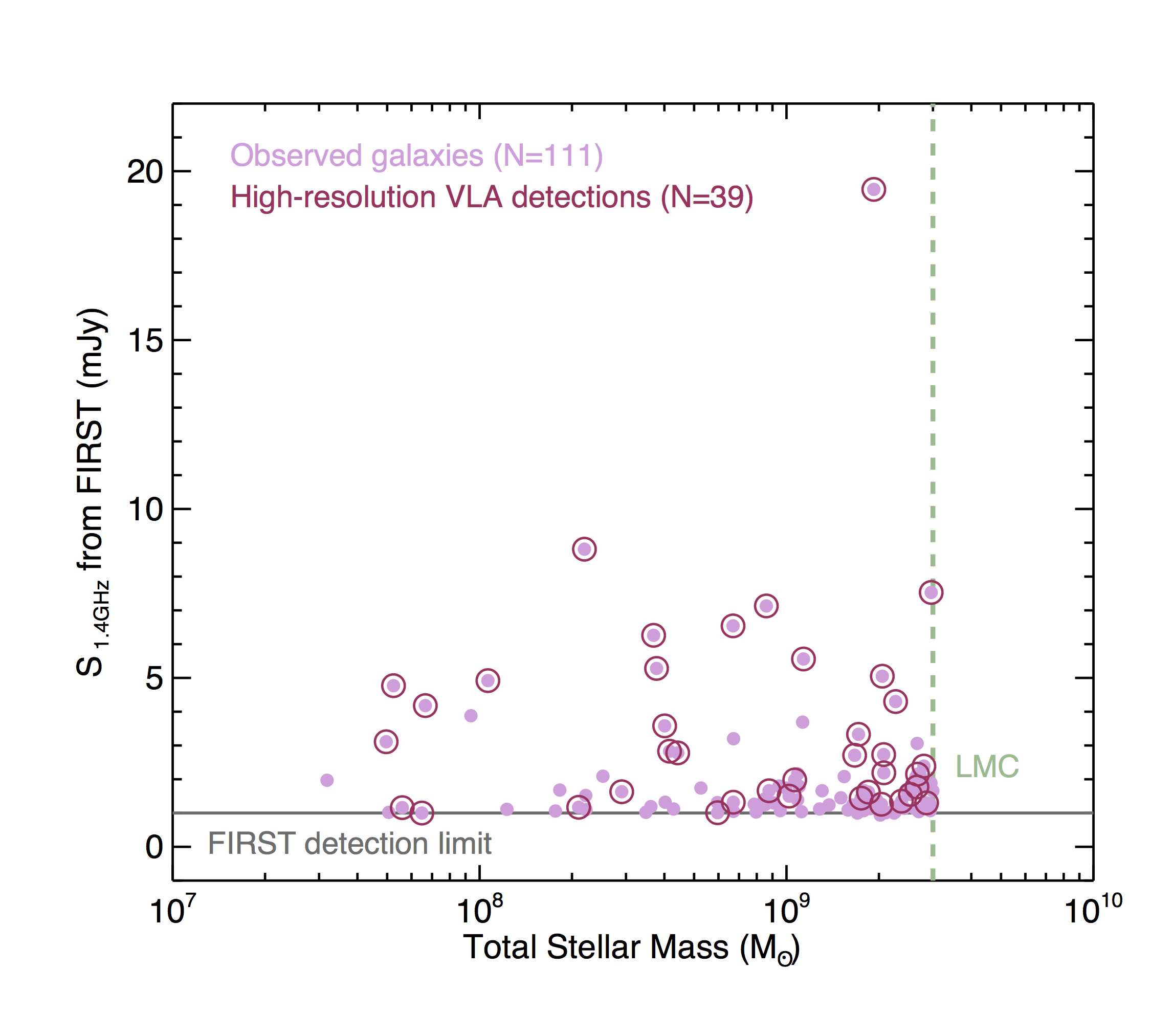} &
\includegraphics[width=3.6in]{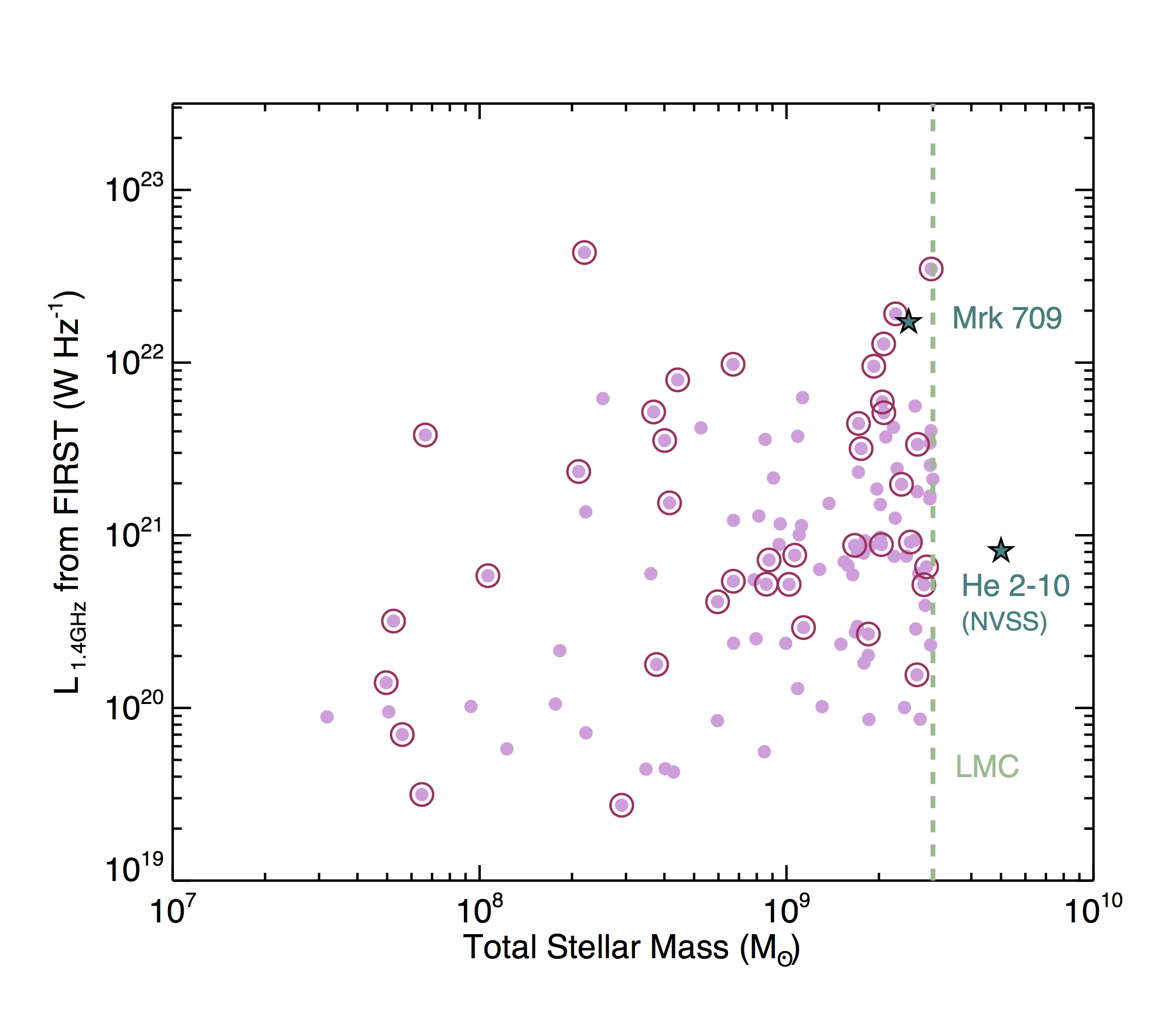} 
\end{array}$
\end{center}
\vspace{-0.75cm}
\caption{Peak flux density (left) and corresponding luminosity density (right) at 1.4 GHz from the FIRST survey versus stellar mass from the NSA for our sample of target galaxies.  We detect compact radio sources with our new VLA observations at 9 GHz towards 39 of the 111 observed galaxies (circled points). The stellar mass of the LMC is indicated as a dashed line. Henize 2-10 \citep{reinesetal2011} and Mrk 709 \citep{reinesetal2014} are shown as teal stars in the right panel.  Henize 2-10 is not in the FIRST footprint so we show the 1.4 GHz NVSS flux density.  Although the NVSS beam is significantly larger than FIRST, the distance to Henize 2-10 is only $\sim 9$ Mpc and the median distance to our VLA sample is $\sim 70$ Mpc.  Therefore the physical resolution is similar.  
\label{fig:firstmass}}
\end{figure*}

There are observational and theoretical arguments for why this strategy is appealing.  First, we know from a number of observational scaling relations that the smallest BHs should live in low-mass galaxies (e.g., \citealt{mcconnellma2013}; \citealt{kormendyho2013}; \citealt{reinesvolonteri2015}; \citealt{bentzetal2018}; \citealt{schutteetal2019}).  Second, models of BH growth in a cosmological context indicate that the signatures of BH seeding should be strongest in dwarf galaxies, with the BH occupation fraction and scaling relations at low masses varying depending on whether the seeds are light or heavy \citep{volonterietal2008,volonterinatarajan2009,vanwassenhoveetal2010,ricartenatarajan2018}.  Recent simulations also suggest that dwarfs should contain relatively pristine BHs due to supernova feedback stunting BH growth \citep{habouzitetal2017,anglesalcazaretal2017}.  Finally, from a practical standpoint, finding and studying active massive BHs with $M_{\rm BH} \lesssim 10^5 M_\odot$ in present-day dwarf galaxies is actually achievable with current capabilities \citep[e.g.,][]{reinesetal2013,baldassareetal2015}.  

In general, there are many ways to identify the presence of an accreting massive BH (see \citealt{ho2008} for a review).  Broad H$\alpha$ emission has been used to identify low-mass AGNs \citep{greeneho2004,greeneho2007,dongetal2012,reinesetal2013} and accompanying narrow-line AGN signatures can be use to firmly rule out supernova interlopers \citep{baldassareetal2016}.  Optical narrow emission-line diagnostic diagrams (e.g., the BPT diagram after \citealt{baldwinetal1981}) are used to separate line-emitting galaxies by their primary excitation source (AGN vs.\ massive stars) and have been applied to the general galaxy population in the Sloan Digital Sky Survey (SDSS) (e.g., \citealt{kauffmannetal2003agn,kewleyetal2006}) and with a focus on low-mass \citep{barthetal2008a} and dwarf galaxies (\citealt{reinesetal2013, moranetal2014, sartorietal2015}).  However, optical diagnostics are only sensitive to BHs radiating at significant fractions of their Eddington luminosity, which are rare compared to lower Eddington-ratio systems \citep{schulzewisotzki2010}.  Furthermore, low-mass galaxies generally have ongoing star formation, gas and dust that can mask or extinguish the optical signatures of BH accretion.  Therefore, while there may be an accreting BH present at the center of a dwarf galaxy, the total observed line emission in the SDSS 3\arcsec\ aperture may be dominated by star formation.  Even without significant ongoing star formation, AGN signatures may be heavily diluted by host galaxy light such that the emission lines are effectively hidden \citep{moranetal2002}.  Therefore, alternative search techniques (e.g., optical variability; \citealt{baldassareetal2018}) and observations at other wavelengths are needed to make further progress on the demographics of massive BHs in dwarf galaxies.

X-ray observations can also be used to search for massive BHs in dwarf galaxies \citep{galloetal2008,schrammetal2013,reinesetal2014, lemonsetal2015, milleretal2015, secrestetal2015, pardoetal2016, reinesetal2016, baldassareetal2017a, chenetal2017, mezcuaetal2018, latimeretal2019}, however long exposure times (hours) are typically necessary to detect low-luminosity AGN even at modest distances. Therefore, it is currently not practical to conduct a large-scale X-ray survey of dwarf galaxies to search for low-mass BHs with correspondingly low luminosities.  Moreover, X-ray observations alone cannot always firmly establish the presence of a massive BH since stellar-mass X-ray binaries can have luminosities comparable to weakly accreting massive BHs. 

Mid-infrared searches, based on {\it WISE} for example \citep{satyapaletal2014, sartorietal2015, marleauetal2017, kavirajetal2019}, are fraught with other problems. In principle very red mid-IR colors can be used to infer the presence of an AGN, however at the low luminosities associated with dwarf galaxies the interpretation is less clear as red mid-IR colors can also be attributed to intense star formation \citep{hainlineetal2016, satyapaletal2018}.  Moreover, the resolution ($6\arcsec-12\arcsec$ for {\it WISE}) is not sufficient to disentangle potential AGN emission from star formation related emission.  The detection of infrared coronal lines with the {\it James Webb Space Telescope} may prove to be a useful diagnostic for probing low-mass BHs \citep{cannetal2018}, however large surveys of dwarf galaxies intended to identify new low-mass AGNs will likely be prohibitively expensive.

Here we undertake a large-scale radio search for massive BHs in dwarf galaxies using high-resolution observations from the NSF's Karl G.\ Jansky Very Large Array (VLA).
This work is motivated in part by the discovery of the first radio (and X-ray) detected massive BH in a dwarf starburst galaxy, Henize 2-10 \citep{reinesetal2011, reinesdeller2012,  reinesetal2016}.  
Furthermore, radio observations with the VLA are ideally suited for a large-scale survey of dwarf galaxies.  First, AGNs (even weak ones) almost always produce radio continuum at centimeter wavelengths (\citealt{ho2008} and references therein), which is virtually immune to dust extinction.  Second, the sub-arcsecond spatial resolution afforded by the VLA in the A-configuration is sufficient to pinpoint the location of any compact radio emission and isolate it from the diffuse synchrotron emission emitted by 
the host galaxy.  Finally, the exquisite sensitivity of the VLA now enables efficient observations of a large number of dwarf galaxies.  This was not possible before the enormous increase in bandwidth provided by the Expanded VLA project.  

\section{Sample of Dwarf Galaxies}\label{sec:sample}

As in \citet{reinesetal2013}, we construct our parent sample of dwarf galaxies from the NASA-Sloan Atlas (NSA; version {\texttt v0\_1\_2}), which is a catalog of images and parameters of SDSS galaxies with redshifts $z < 0.055~(D \lesssim 225$~Mpc).  {Redshifts primarily come from SDSS DR8 spectra. However, in some cases values are taken from other sources\footnote{http://www.nsatlas.org/documentation} including the NASA Extragalactic Database, the Six-degree Field Galaxy Redshift Survey, the Two-degree Field Galaxy Redshift Survey, the CfA Redshift Survey (ZCAT) and ALFALFA.} Stellar masses are provided in the NSA and are derived from the {\texttt kcorrect} code of \citet{blantonroweis2007}, which fits broadband {\it GALEX} and SDSS fluxes using templates based on the stellar population synthesis models of \citet{bruzualcharlot2003} and the nebular emission-line models of \citet{kewleyetal2001}.  

We first select sources with stellar masses $M_\star \leq 3 \times 10^9~M_\odot$ in the NSA, as well as impose modest absolute magnitude cuts of $M_g$ and $M_r > -20$ to help mitigate selecting luminous/massive galaxies with erroneous mass estimates.  Following Reines et al.\ (2013), our upper mass limit is approximately equal to the stellar mass of the Large Magellanic Cloud (LMC), which is the most massive dwarf satellite of the Milky Way.  The minimum stellar mass of our sample is $M_\star \gtrsim 10^7~M_\odot$ and is due to the SDSS spectroscopic apparent magnitude limit of r $<$ 17.7. Our parent sample includes 43,707 objects.

To maximize the number of detections and to target galaxies with radio luminosities similar to Henize 2-10, the prototypical dwarf galaxy with a radio-selected massive BH \citep{reinesetal2011, reinesdeller2012}, we select a sub-sample of dwarf galaxies detected in the VLA Faint Images of the Radio Sky at Twenty-centimeters (FIRST) Survey \citep{beckeretal1995}.  We cross-correlate our parent sample of 43,707 dwarfs to the FIRST catalog (version 2013Jun5) requiring a match radius $\leq 5$\arcsec ($\sim$ the resolution of FIRST), and find 186 matches throughout the NSA volume.  We then visually inspect all of the matches and eliminate 35 {interlopers from the sample that are clearly not dwarf galaxies} (e.g., nearby H{\footnotesize II} regions and distant quasars) and 3 nearby, well-studied dwarf galaxies with archival VLA data (including NGC 4395 that is already known to host an AGN), leaving us with 148 objects.  Of these, 111 dwarf galaxies with FIRST detections were observed with the VLA for this program (see Figure \ref{fig:firstmass} and Table \ref{tab:sample}).  The remaining galaxies were not observed due to scheduling priorities.  

Our radio-selected dwarf galaxies are quite rare -- only $\sim$0.3\% of sources in our parent sample of dwarf galaxies have counterparts in FIRST.  However, we note that many AGNs, even in low-redshift galaxies, do not produce radio continuum emission that is detectable at FIRST sensitivity levels \citep{condonetal2019} and therefore we could be missing a significant population of massive BHs.  While our sample of dwarf galaxies is clearly biased, targeting a radio-selected sample virtually ensures there is either an AGN, intense star formation, or both.  Even if the majority of the radio emission detected in FIRST is due to star formation, the sensitivity and high angular resolution of our VLA observations will enable the detection of massive BHs in our target dwarf galaxies, so long as they are present and accreting at modest rates. 

\subsection{Potential Contamination from Background Radio Sources}\label{sec:background}

\begin{figure}[!t]
\begin{center}
\hspace{-.35cm}
\includegraphics[width=3.5in]{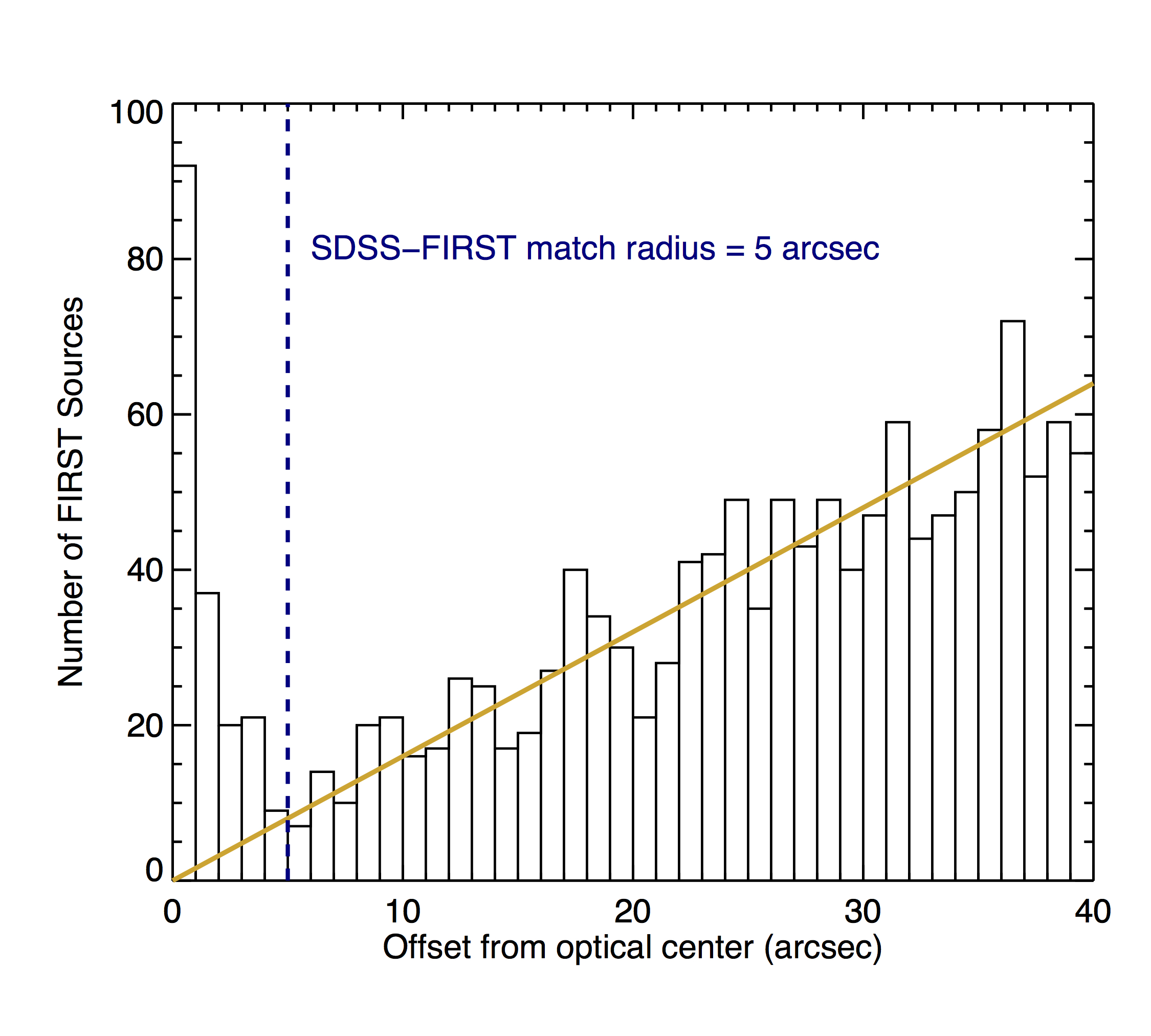}
\end{center}
\vspace{-0.7cm}
\caption{Observed offset distribution from cross-matching our SDSS parent sample with the FIRST survey using a match radius of 40\arcsec.  We select our target galaxies to have offsets $\leq 5$\arcsec\ (blue line).  The yellow line shows the expected number of chance alignments with background sources as a function of offset between an SDSS and FIRST source.
\label{fig:offsethist}}
\end{figure}

We assess the likelihood of chance alignments between galaxies in our parent sample and background radio sources using two methods.  First we estimate $N(S_{\rm min})$, the cumulative number of radio source counts per steradian with 1.4 GHz flux densities greater than $S_{\rm min}$, using the results of \citet{condon1984}.  Taking $S_{\rm min} = 1.5~{\rm mJy}$ (i.e., the completeness limit of FIRST) and multiplying $N(S_{\rm min}) \approx 219,000$ str$^{-1}$ by the area enclosed by a 5\arcsec\ circle gives the expected number of background sources for a given galaxy, $N_{\rm bk,gal} = 0.0004$.  Across our entire parent sample of 43,707 galaxies, we expect $N_{\rm bk,samp} \approx 18 \pm \sqrt{18} \approx 18 \pm 4$ background sources.  

We also estimate the expected number of background sources using the data in Figure \ref{fig:offsethist}, which shows the observed offset distribution from cross-matching our SDSS sample and FIRST using a match radius of 40\arcsec.  The offset probability histogram for background sources should be a Rayleigh distribution, which equals zero at an offset of zero and rises linearly for small offsets (see the yellow line in Figure \ref{fig:offsethist}).  The observed distribution is a minimum at our adopted match radius of 5\arcsec.  At larger offsets the number of sources per arcsecond of offset, $N(d_{\rm off})$, is well-fit by $N(d_{\rm off}) = 1.6~d_{\rm off}$, where the offset $d_{\rm off}$ is in units of arcsec.  The total estimated number of background sources with offsets less than 5\arcsec\ is found by integrating $N(d_{\rm off})$ from $d_{\rm off}$ =  0\arcsec\ to 5\arcsec, which gives $N_{\rm bk,samp} \approx 20 \pm \sqrt{20} \approx 20 \pm 4$ sources.  This is consistent with our calculation above using known radio source counts, and we adopt the average of $N_{\rm bk,samp} \approx 19 \pm 4$.

The $19 \pm 4$ chance alignments with background radio sources are expected to be present in the 186 initial matches between our parent sample of 43,707 SDSS objects and FIRST radio sources (using a match radius of 5\arcsec).  Given that we obtained new VLA observations for only 111 galaxies of the 186 matches, we expect $111/186 \times 19 \approx 11 \pm 3$ background sources to be present in our observed sample (i.e., roughly 10\% contamination).  The background contaminants will be dominated by AGNs/quasars at FIRST flux densities $S_{1.4 \rm GHz} \gtrsim 1$ mJy \citep[e.g., see Figure 2 in][]{beckeretal1995}.  

\begin{figure}[!t]
\begin{center}
\includegraphics[width=3.4in]{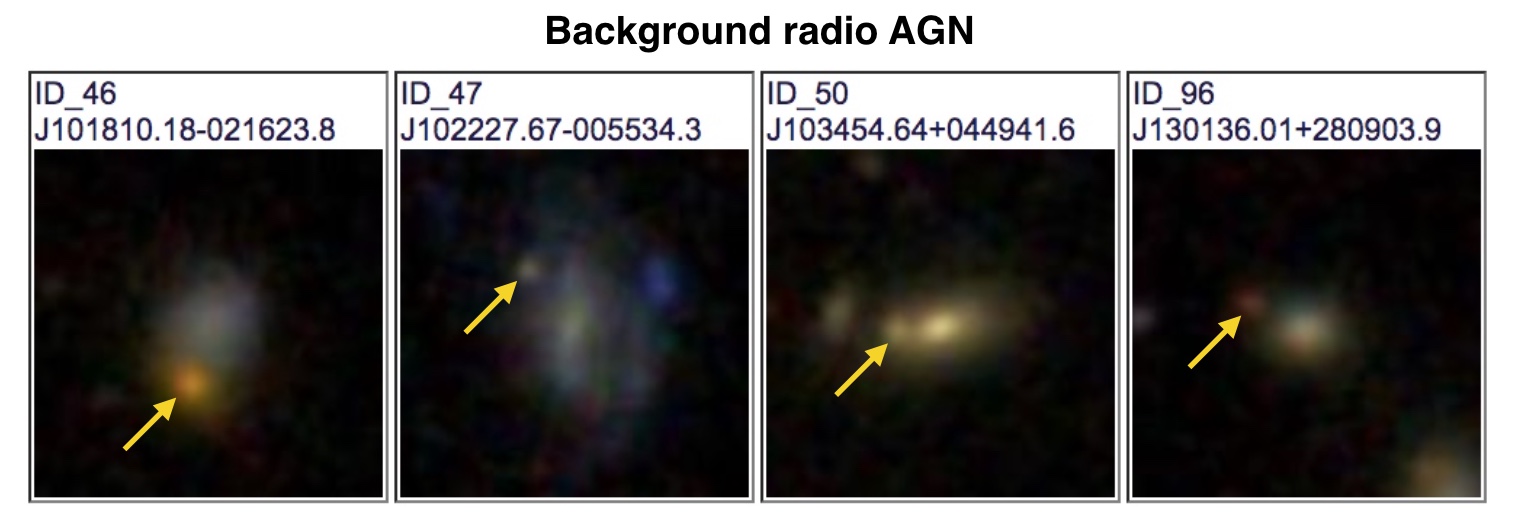}
\end{center}
\vspace{-0.25cm}
\caption{Identifiable background radio contaminants are found in 4 of our 39 galaxies with 9 GHz VLA detections.  The yellow arrows indicate the position of the compact radio emission.  The radio sources have clear optical counterparts and are offset by 2\farcs8 to 4\farcs9 from the optical centers of the target galaxies.  Images are 25\arcsec $\times$ 25\arcsec.  See the Appendix for a discussion of ID 47, which is a dwarf galaxy hosting an {\it optical} AGN.
\label{fig:sdss_backim}}
\end{figure}

It is not immediately clear whether or not we would detect these background AGNs with our new high-resolution VLA observations at 9 GHz.  If the FIRST flux density is dominated by a compact flat spectrum core, then we expect we would.  However, if the FIRST flux density at 5\arcsec\ scales is dominated by more extended steep spectrum lobes, we may not detect the source at high ($\sim 0\farcs25$) resolution.  Therefore we consider the $11 \pm 3$ expected background sources in our observed sample to be an upper limit on the number of background sources in our detected sample.  As discussed in Section \ref{sec:radiosources}, we detect 9 GHz compact radio emission towards 39 of the 111 observed galaxies. {We find 2 confirmed background sources and 2 additional strong background candidates (see Figure \ref{fig:sdss_backim} and footnotes 2-5),} leaving no more than $\sim 7$ contaminants left in the remaining 35 galaxies.   

\section{Observations and Data Reduction}\label{sec:obs}

\begin{figure*}[!t]
\begin{center}
\includegraphics[width=7in]{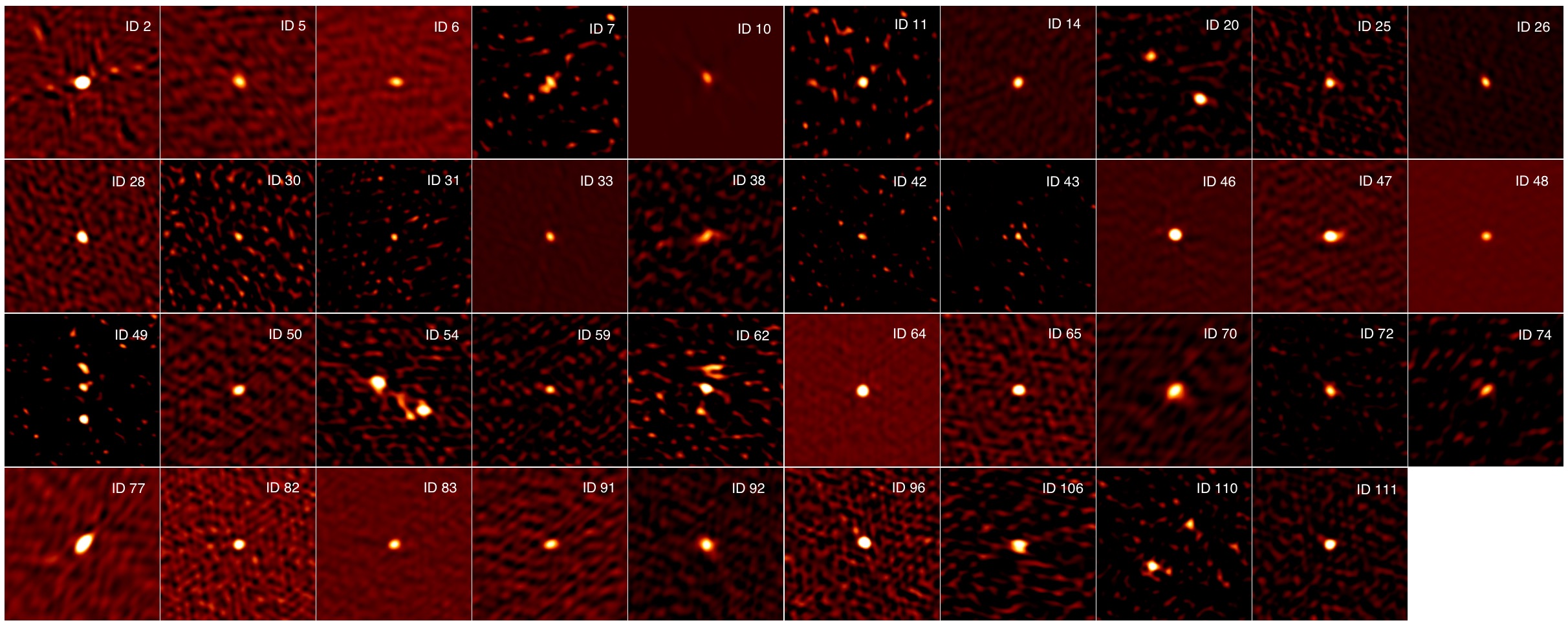}
\end{center}
\caption{9 GHz VLA images of the compact radio sources detected towards our target galaxies, centered on the radio emission.  These images include the objects in Sample A, Sample B, as well as the likely background contaminants towards the galaxies in Figure \ref{fig:sdss_backim}.  All sources are within 5\arcsec\ of the optical center of the galaxy.  Images are $4\arcsec \times 4\arcsec$ and the typical synthesized beam size is $\sim 0\farcs25$. See Table \ref{tab:radio}
for source properties.  
\label{fig:vla_ims}}
\end{figure*}

Our 111 target dwarf galaxies were observed with the VLA in the most extended A-configuration in 2014 from Feb 20 to May 31 (PI: Reines; Project 14A-220).  Continuum observations were carried out at X Band with 2 $\times$ 2 GHz basebands centered at 9.0 and 10.65 GHz.  These observing parameters yield a resolution of $\sim 0\farcs25$ and therefore the beam solid angle is $\sim 400\times$ smaller than FIRST.  At the median distance of our sample of dwarf galaxies ($\sim$70 Mpc), this angular resolution corresponds to a physical scale of $\sim$85 pc.  

We grouped the observations into 1 hr-long scheduling blocks to facilitate dynamic scheduling, typically with five target galaxies per group. On-source integration times were $\sim$ 6--7 minutes per galaxy yielding a typical RMS noise of $\sim15~\mu$Jy beam$^{-1}$ ($\sim10\times$ more sensitive than FIRST).  Nearby phase calibrators were observed for $\sim 1.5$ min before and after each target.  A flux calibrator was also observed during each block. The absolute flux calibration of the VLA data is expected to be $\sim 5$\%. 

The raw data were reduced using CASA version 4.7.2 on the NRAO New Mexico Array Science Center computer cluster.  We first ran the VLA calibration pipeline on each of our 23 scheduling blocks.  Additional flagging was then performed upon inspection of the data and the pipeline was re-run with the additional flags pre-applied.  The final calibrated science target data were split off into the two basebands and imaged separately using multi-frequency synthesis and Briggs weighting.  A summary of our VLA observations is given in Table \ref{tab:vla}.

\section{Compact Radio Sources}\label{sec:radiosources}

We detect a total of 48 compact radio sources towards 39 of our 111 target galaxies (i.e., a 35\% detection rate) with a detection threshold of $\gtrsim 3\sigma$.  {We use the term ``compact" radio sources to differentiate between our VLA detections at high resolution and the radio sources detected towards our target galaxies at lower-resolution in the FIRST survey.}  
 The VLA images are shown in Figure \ref{fig:vla_ims}.  We identify 4 radio detections that appear to be unrelated background objects with point-like optical counterparts (IDs 46\footnote{{We confirm a background radio source towards ID 46. An SDSS/BOSS spectrum of the offset source indicated by the arrow in Figure \ref{fig:sdss_backim} gives a redshift of $z=0.366$.}}, 47\footnote{{We confirm that the offset radio source towards ID 47 is a background AGN.  See the Appendix.}}, 50\footnote{An SDSS spectrum of the target galaxy ID 50 centered in Figure \ref{fig:sdss_backim} indicates a redshift of $z=0.102$, and therefore it is unlikely a dwarf galaxy.  We do not know the redshift of the offset source.}, 96\footnote{The redshift of the target galaxy ID 96 and the offset source are unknown.  Given the large offset and clear optical counterpart of the radio source, the radio source is likely unrelated to the target galaxy.}; see Figure \ref{fig:sdss_backim} and the Appendix).  {For the remaining 35 galaxies with VLA detections, we visually inspect optical or \HI\ spectra (see \S\ref{sec:sample}) to verify redshifts and hence stellar masses.  We are able to confirm that 28 have reliable redshifts and are bona fide dwarf galaxies.  The remaining 7 galaxies have unreliable redshifts due to poor quality spectra and/or photometric redshifts with relatively large uncertainties.}  We refer to the restricted sample of 28 confirmed dwarf galaxies (with 35 compact radio sources) as Sample A, and the remaining 7 galaxies (with 9 compact radio sources) as Sample B (see Table \ref{tab:radio}). 

Radio flux densities at 9.0 and 10.65 GHz are measured interactively in the CASA viewer using two different methods.  First, we attempt to fit the detected sources with a single Gaussian model using the task IMFIT.  This provides a good fit for the majority of sources.  At the resolution of our observations ($\sim$ 0\farcs25), many of the detections are consistent with being point sources (Table \ref{tab:radio}) {as determined from IMFIT}.  We also measure flux densities with a variety of user-defined apertures that trace the shape of the source.  This is particularly useful for those sources with extended or irregularly-shaped emission profiles. 

\begin{figure*}[!t]
\begin{center}$
\begin{array}{cc}
\hspace{-0.25cm}
\includegraphics[width=3.3in]{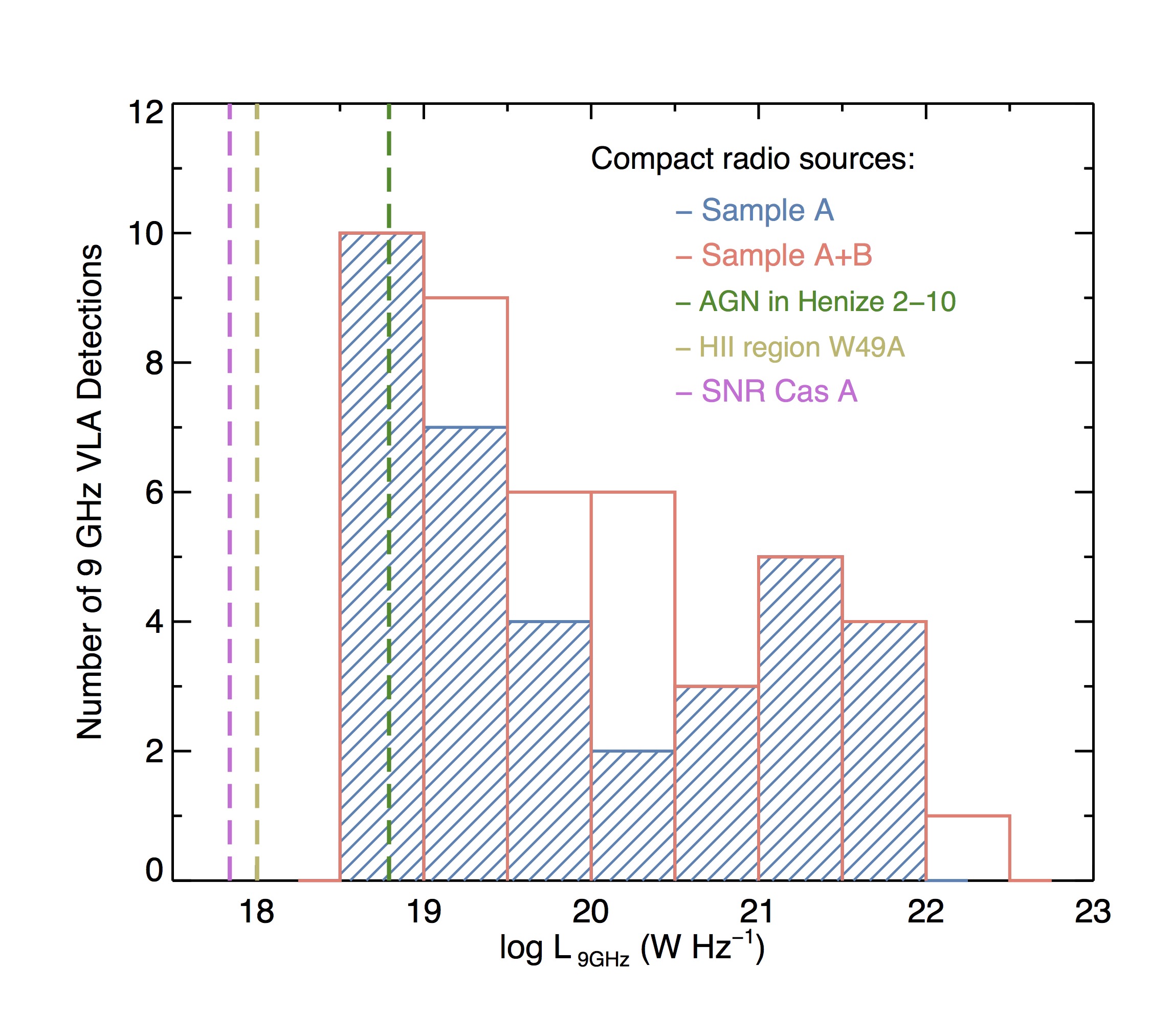} &
{\includegraphics[width=3.3in]{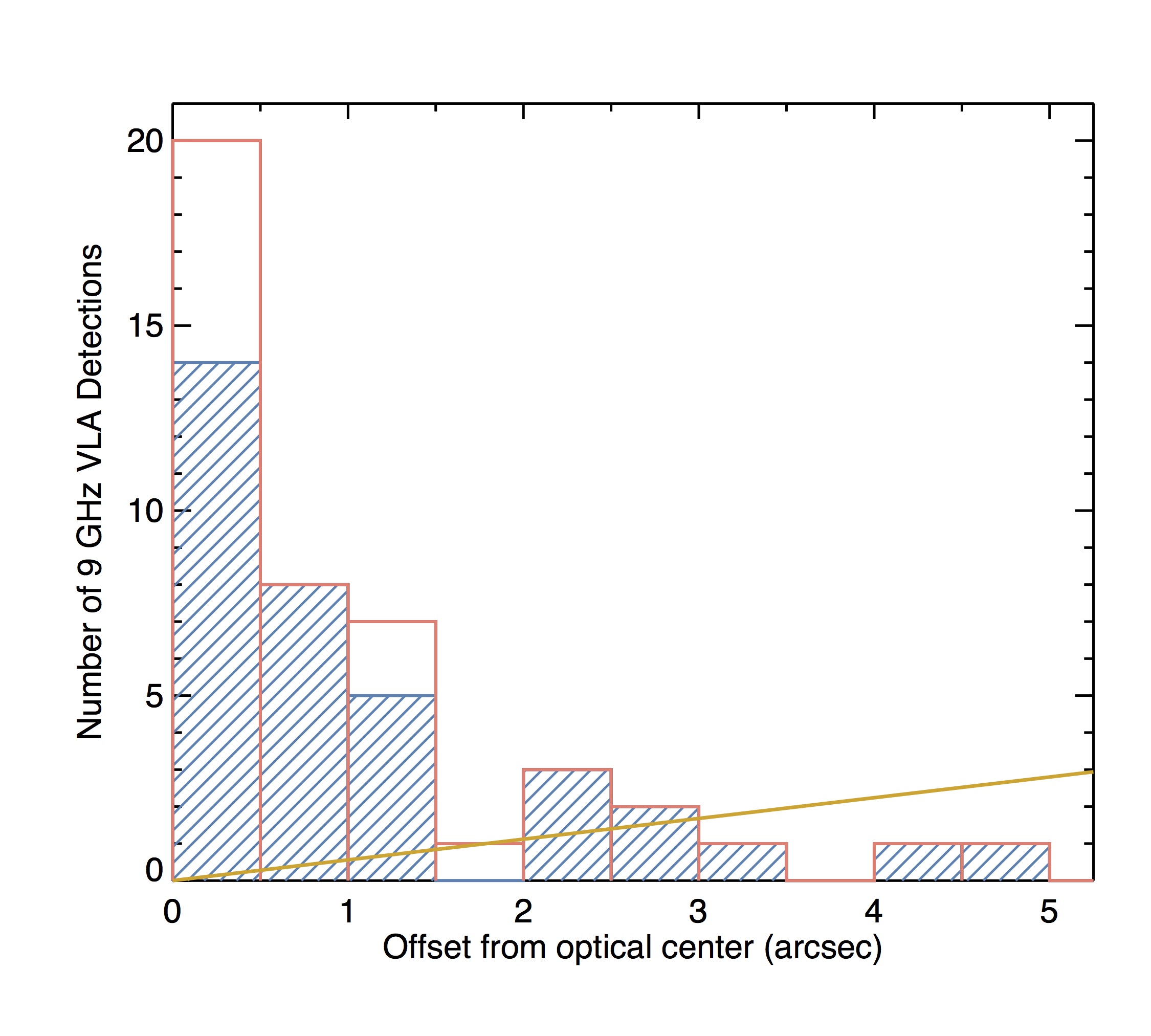}}
\end{array}$
\end{center}
\vspace{-0.75cm}
\caption{Left: Distribution of the 9 GHz spectral luminosities of the compact radio sources detected in our VLA observations.  For comparison, we show the values of the low-luminosity AGN in Henize 2-10, the young Galactic supernova remnant Cas A, and the young Galactic \HII\ region W49A.
Right: Distribution of the offsets between the positions of detected compact radio sources and optical positions of the target galaxies (see Table \ref{tab:radio}).  The yellow line shows the expected distribution of background radio sources under the most conservative assumptions discussed in Sections \ref{sec:background} and \ref{sec:radiosources}.
\label{fig:Ldet}}
\end{figure*}

Table \ref{tab:radio} lists properties of the detected compact radio sources including coordinates, flux densities, luminosity densities, spectral indices, and radio-optical positional offsets.  The 9 GHz spectral luminosities in Sample A are in the range $L_{\rm 9 GHz} \sim 10^{18.5}-10^{22.0}$ W Hz$^{-1}$ with a median value of $10^{19.6}$ W Hz$^{-1}$ (see Figure \ref{fig:Ldet}).  For comparison, the low-luminosity AGN in Henize 2-10 has a spectral luminosity of $\sim 10^{19}$ W Hz$^{-1}$.  We calculate spectral indices, $\alpha$, between 9.0 and 10.65 GHz using the measured flux densities in each baseband ($S_\nu \propto \nu^\alpha$).  Given the small spread in frequency and typical signal-to-noise ratios, the uncertainties are rather large except for the most luminous sources (Table \ref{tab:radio}).  We detect multiple compact radio sources in 3 galaxies in Sample A and 2 additional galaxies in Sample B.  

By design, all of the radio sources are within 5\arcsec\ of the optical center of the galaxy as defined by the SDSS images.   
The observed distribution of radio-optical positional offsets for all 44 detections (after removing identifiable background contaminants) is shown in the right panel of Figure \ref{fig:Ldet}.  {We expect at most $7 \pm 3$ remaining background radio contaminants in Samples A and B}, where the probability of contamination rises linearly with separation between the radio source and the optical center of the galaxy.  We construct the corresponding offset distribution of possible background contaminants by scaling the relation shown in Figure \ref{fig:offsethist} such that $N(d_{\rm off}) = A \times 1.6~d_{\rm off}$, where $A$ is the scale factor and $d_{\rm off}$ is the offset in arcsec.  Integrating between 0\arcsec\ and 5\arcsec\ and setting the result equal to 7 sources gives $A=0.35$.  The resulting distribution of possible background sources, which we consider an upper limit, is shown in Figure \ref{fig:Ldet}.     

We do not detect radio emission with our new high-resolution VLA observations towards 72 of the 111 target galaxies with FIRST detections.  In these cases, the FIRST sources are likely dominated by more extended and diffuse radio emission from star formation processes (or possibly large scale AGN radio lobes), and may include some fraction of the expected background contaminants.  While we cannot rule out the presence of massive BHs in these 72 galaxies, we have no evidence for their existence.

\section{Origin of the Compact Radio Emission}\label{sec:origin}

Our primary goal is to detect compact radio emission from accreting massive BHs in our target galaxies.  However, at centimeter radio wavelengths, we need to consider possible contamination in our sample from supernova remnants (SNRs), young supernovae (SNe), and ionized gas from \HII\ regions.  Stellar-mass X-ray binaries are not a concern as they do not exceed $\sim 10^{17}$ W Hz$^{-1}$ in the radio even when flaring \citep{corbeletal2012,galloetal2018}.  Out of 44 detected compact radio sources (see Section \ref{sec:radiosources}), we expect at most $\sim 7 \pm 3$ background contaminants that are primarily at large ($\gtrsim$ 2\arcsec) radio-optical positional offsets (Figure \ref{fig:Ldet}; Sections \ref{sec:background} and \ref{sec:radiosources}). As demonstrated below, at least 13 compact radio sources in Sample A are almost certainly due to AGN activity.  Another 6 (likely more massive) galaxies in Sample B also have detectable radio AGNs, with one of these systems hosting a possible dual AGN.

\subsection{Thermal \HII\ Regions?}\label{sec:HII}

We begin by considering whether the detected compact radio emission could be due to thermal bremsstrahlung (i.e., free-free emission) from ionized hydrogen in star-forming regions.  Under the assumption that the detected compact radio emission is thermal bremsstrahlung from \HII\ regions, we can estimate the production rate of Lyman continuum (i.e., ionizing) photons, which in turn can be used to estimate the stellar content and instantaneous star formation rates (SFRs) in the regions.
The production rate of Lyman continuum photons, $Q_{\rm Lyc}$, is given by \citet{condon1992}:

\begin{eqnarray}
\left({Q_{\rm Lyc} \over {\rm s^{-1}}}\right) \gtrsim 6.3\times10^{52} \left({T_e \over 10^4{\rm ~K}}\right)^{-0.45}
\left({\nu \over {\rm GHz}}\right)^{0.1}  \nonumber \\
\times \left({L_{\nu, \rm thermal} \over 10^{20} {\rm ~W ~Hz^{-1}}}\right),
\label{Qlyc}
\end{eqnarray}

\noindent
{where the inequality allows for dust absorption of ionizing photons.} 
Using the 9 GHz spectral luminosities of our sources in Sample A (\S \ref{sec:radiosources}) for $L_{\nu, \rm thermal}$ and adopting an electron temperature of $T_e = 10^4$~K, the ionizing luminosities would be in the range log $Q_{\rm Lyc} \sim 51.4 - 54.9$ (median value of log $Q_{\rm Lyc} \sim 52.5$) under the assumption that the radio emission is thermal.  
We estimate the number of massive stars that would be necessary to power this ionizing radiation using the results from  \citet{vaccaetal1996};  a ``typical" O-type star (type O7.5 V) produces $Q_{\rm Lyc} = 10^{49}$ s$^{-1}$.  Therefore, approximately $260$ to $7.5 \times 10^5$ typical O-type stars would be necessary to account for the observed radio emission in our sources.  

For comparison, we consider the Galactic thermal radio source W49A \citep{mezgeretal1967}, which is one of the youngest and most luminous star-forming regions in the Milky Way, as well as known extragalactic young massive star clusters.  We calculate a 9 GHz luminosity density of $L_{\rm 9 GHz} = 10^{18.0}$ W Hz$^{-1}$ for W49A using a flux density of 65 Jy from \citet{mezgeretal1967} and a distance of 11.4 kpc from \citet{gwinnetal1992}.  Using Equation \ref{Qlyc} yields an ionizing luminosity of log $Q_{\rm Lyc} \sim 50.9$, or the equivalent of $79$ typical O-type stars. Figure \ref{fig:Ldet} (left panel) shows that the compact radio sources in our sample are orders of magnitude more luminous than the Galactic massive star-forming region W49A.  \HII\ regions produced by extragalactic young massive star clusters can also be identified as compact free-free (thermal) radio sources \citep[e.g.,][]{kobulnickyjohnson1999, turneretal2000,reinesetal2008a,johnsonetal2009,aversaetal2011,kepleyetal2014}, yet even the most extreme sources have log $Q_{\rm Lyc} \lesssim 10^{53}$ \citep{turneretal2000,johnsonetal2009}.  

We estimate the instantaneous SFRs from the ionizing luminosities calculated above for our sample following \citet{kennicutt1998}:

\begin{equation}
{\rm SFR (}M_\odot~{\rm year}^{-1}) = 1.08 \times 10^{-53}~Q_{\rm Lyc} ({\rm s}^{-1})
\end{equation}

\noindent
and compare these results to the star formation rates (SFRs) of the entire host galaxies derived from {\it Galaxy Evolution Explorer (GALEX)} far-ultraviolet (FUV) magnitudes from the NSA and {\it Wide-field Infrared Survey Explorer} ({\it WISE}) 22 $\mu$m data from the AllWISE Source Catalog.  We calculate the host galaxy dust-corrected SFRs following \citet{kennicuttevans2012} and \citet{Haoetal2011}:

\begin{equation}
{\rm log~SFR = log}~L{\rm (FUV)_{corr} - 43.35}
\label{eqn:sf1}
\end{equation}

\noindent
where

\begin{equation}
L{\rm (FUV)_{corr}} = L{\rm (FUV)_{obs}} + 3.89~L{\rm (25\mu m)}
\label{eqn:sf2}
\end{equation}

\noindent
and luminosities are in units of erg s$^{-1}$.  {The $1\sigma$ uncertainty using this method is $\sim$0.13 dex \citep{Haoetal2011}.}  We estimate $L$(25 $\mu$m) from the {\it WISE} 22 $\mu$m data since the flux density ratio for these bands is expected to be of order unity for both late-type and early-type galaxies \citep{jarrettetal2013}.  Nine of the galaxies are not detected at 22 $\mu$m; in these cases we derive SFRs using the uncorrected observed FUV luminosities.  

\begin{figure*}[!t]
\begin{center}$
\begin{array}{ccc}
\includegraphics[width=2.35in]{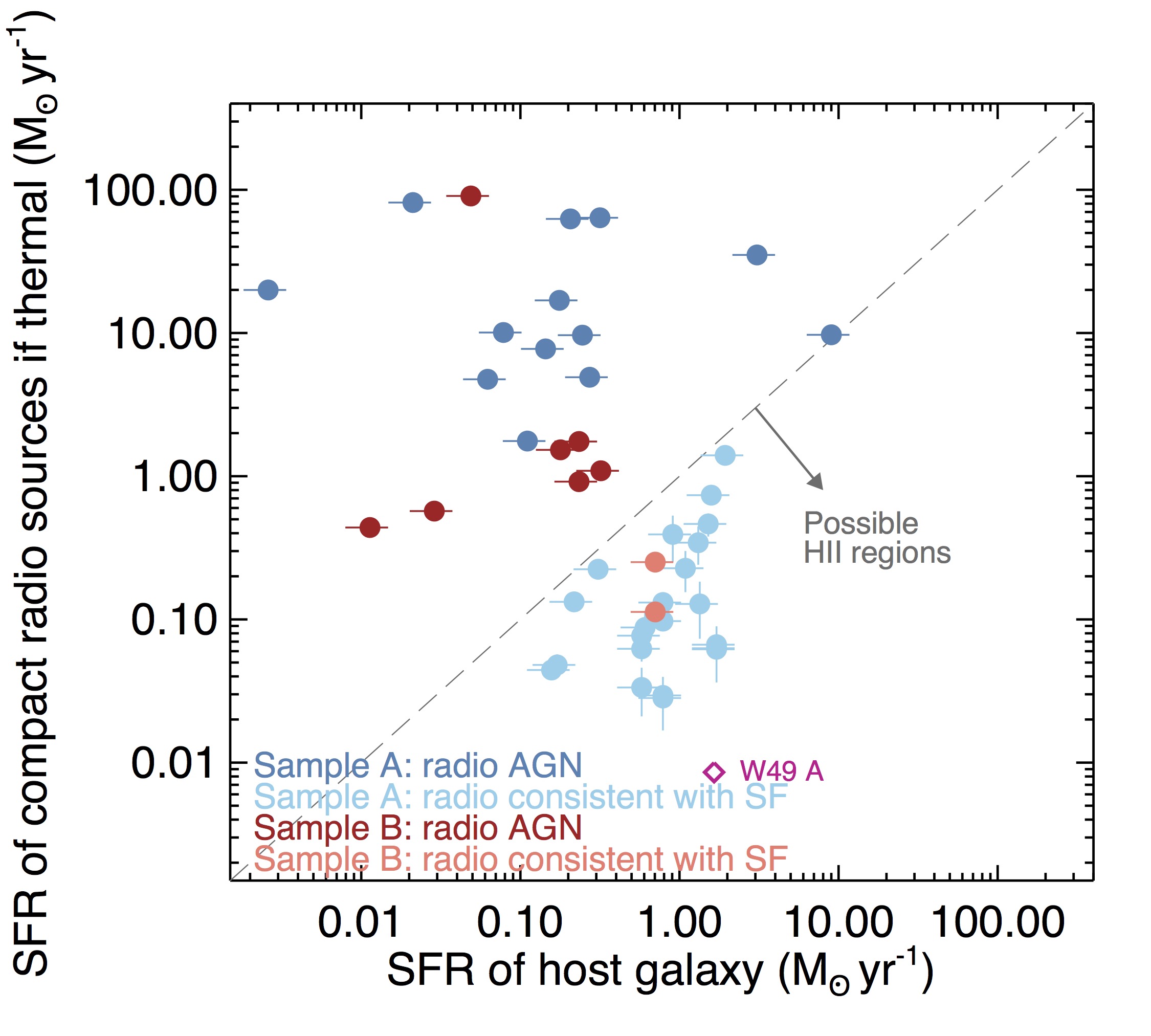} &
\hspace{-.25cm}
\includegraphics[width=2.35in]{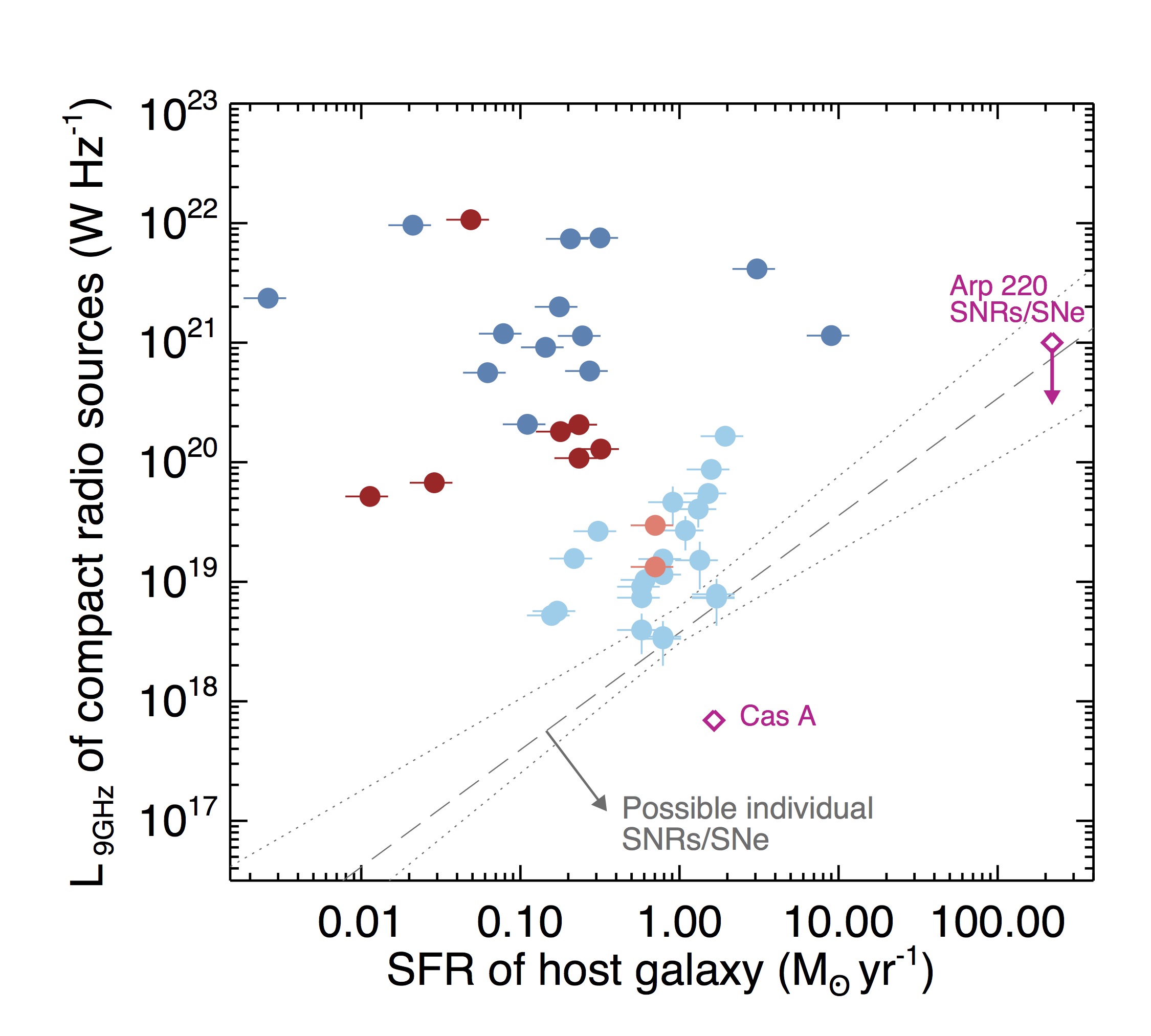} &
\hspace{-.25cm}
\includegraphics[width=2.35in]{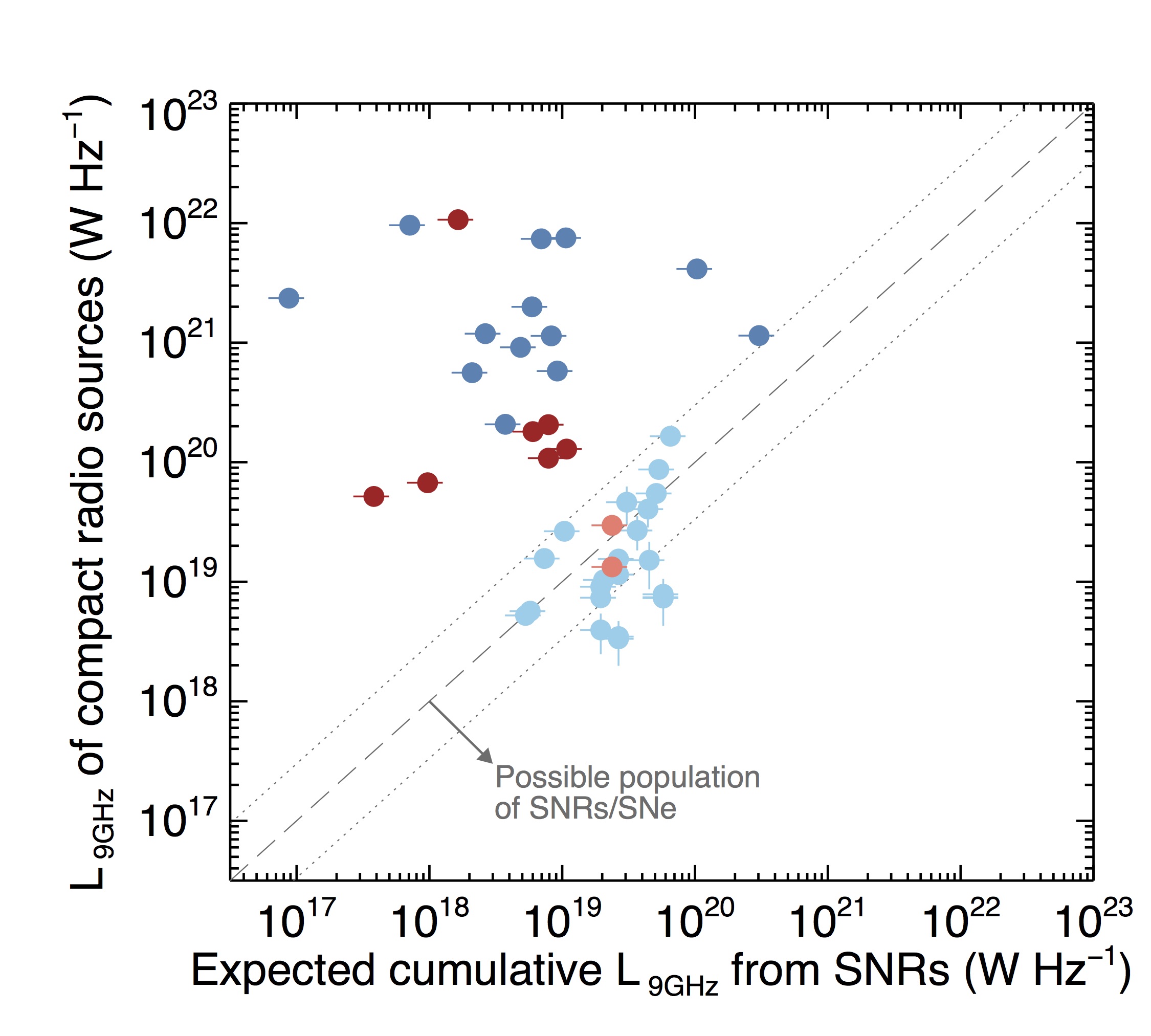}
\end{array}$
\end{center}
\vspace{-0.3cm}
\caption{Possible origins of the compact radio sources detected in our sample of dwarf galaxies.
{Left}: Under the assumption that the radio emission is thermal bremsstrahlung (i.e., free-free emission from ionized gas in star-forming regions), we plot the instantaneous SFRs of the compact radio sources versus the SFRs of the entire galaxies (\S\ref{sec:HII}).  Sources above the one-to-one relation cannot be explained by this mechanism since individual star-forming regions cannot have SFRs exceeding the entire galaxy.  Sources below the relation can be, but are not necessarily, \HII\ regions.  The Galactic massive star-forming region W49A is shown for comparison where the SFR of the MW is taken from \citet{licquianewman2015}. {Unseen error bars are smaller than the size of the plotted symbol.}
{Middle}: 9 GHz luminosity densities of the compact radio sources versus SFR of the host galaxies.  The dashed line shows the relation between the most luminous SNR in a galaxy versus SFR of the galaxy from \citet{chomiukwilcots2009}, with dotted lines representing the expected scatter due to random statistical sampling (\S\ref{sec:sn}).  Nearly all of our detected compact radio sources are above the relation and therefore too luminous to be individual SNRs.  The Galactic SNR Cas A and Arp 220 are shown for comparison.  Arp 220 has a SFR of $\sim 220~M_\odot$ yr$^{-1}$ \citep{vareniusetal2016} and all of the radio SNRs/SNe have $L_{\rm 9 GHz} \lesssim L_{\rm 5 GHz} < 10^{21}$ W Hz$^{-1}$ \citep{vareniusetal2019}.
{Right:} 9 GHz luminosity densities of the compact radio sources versus the expected cumulative luminosity from a population of SNRs/SNe in the host galaxies (\S\ref{sec:snpop}).  The dashed line indicates the one-to-one relation and the dotted lines are offset by a factor of three.
\label{fig:sne}}
\end{figure*}

Under the assumption that the compact radio emission is thermal, 13 of the 35 compact radio sources in Sample A would have SFRs larger than the SFRs of their entire host galaxies (see left panel of Figure \ref{fig:sne})\footnote{We note that the dark blue point towards the right (ID 92) is consistent with falling on the one-to-one relation when uncertainties are taken into account.}.  Clearly this is highly unlikely, enabling us to reasonably rule out thermal bremsstrahlung as the origin of the radio emission for these objects.  The calculated SFRs (assuming the compact radio emission is thermal) are $\sim 2$ to 81 $M_\odot$ yr$^{-1}$, which is up to two orders of magnitude higher than the SFRs of well-studied extragalactic \HII\ regions as noted above.  

The implied SFR surface densities, $\Sigma_{\rm SFR}$, of these 13 compact radio sources (if thermal) are also extraordinarily high, ranging from $\Sigma_{\rm SFR} \sim 320 - 4100~M_\odot$~yr$^{-1}$ kpc$^{-2}$.  All of the sources would have $\Sigma_{\rm SFR}$ greater than the ``starburst intensity limit" of 45 $M_\odot$ yr$^{-1}$ kpc$^{-2}$ proposed by \citet{meureretal1997}. They would also be higher than all 61 of the normal spiral galaxies and all but one (Arp 220) of the 36 IR-selected circumnuclear starburst galaxies in the compilation of \citet{kennicutt1998}.  Given all of the findings above, it is highly unlikely that star-forming regions producing free-free radiation can fully account for the compact radio sources detected in our sample of dwarf galaxies.

\subsection{Individual SNRs/SNe?}\label{sec:sn}

We also consider whether individual SNRs or younger SNe could plausibly account for the compact radio emission observed in our sample of dwarf galaxies.  First we compare the 9 GHz spectral luminosities of our detected sources with that of Cas A, which is one of the youngest and most luminous SNRs in the Milky Way with an age of $\sim 300$ years.  Adopting a 1 GHz flux density of $S_{\rm 1 GHz}=2723$ Jy and a spectral index of $\alpha = -0.770$ from \citet{baarsetal1977} (epoch 1980.0), and distance of 3.4 kpc from \citet{reedetal1995}, we calculate a 9 GHz spectral luminosity of  $L_{\rm 9 GHz} \sim 7 \times 10^{17}$ W Hz$^{-1}$ for Cas A.  Figure \ref{fig:Ldet} shows that our detected compact radio sources are orders of magnitude more luminous than Cas A. 

Younger radio supernovae (SNe) have typical ages less than a few decades.  These radio sources can be significantly more luminous than older SNRs, albeit for a relatively short amount of time.  During the SN stage, the blast wave from the explosion is thought to expand in a dense circumstellar medium, whereas in the SNR stage, the blast wave is thought to interact with the surrounding interstellar medium.
\citet{vareniusetal2019} and \citet{ulvestad2009} study the populations of radio SNe/SNRs in the luminous infrared merging galaxies Arp 220 and Arp 299, respectively.  The 5 GHz luminosity densities of 88 sources in Arp 220 are shown in Figure 6 of  \citet{vareniusetal2019} and the 8.4 GHz luminosity densities of 14 sources in Arp 299 are given in Figure 7 of \citet{ulvestad2009}.  The vast majority of the radio SNe in these works have luminosity densities $L_\nu \lesssim 10^{20}$ W Hz$^{-1}$, with the brightest sources having $L_\nu < 10^{21}$ W Hz$^{-1}$.  We consider this an upper limit on $L_{\rm 9 GHz}$ for these sources since SNe/SNRs typically have steep radio spectra that decrease with increasing frequency.   Nine of the compact radio sources in our Sample A have luminosity densities higher than all of $\sim100$ SNe/SNRs found in Arp 220 and Arp 299.  

To more robustly determine if individual SNRs/SNe could produce the observed radio emission, we make use of the relation between the luminosity density of the brightest SNR/SNe in a galaxy and the SFR of the galaxy given by \citet{chomiukwilcots2009}:

\begin{equation}
L_{1.4}^{\rm max} = (95^{+31}_{-23}) {\rm SFR}^{0.98 \pm 0.12}
\end{equation}

\noindent
where the 1.4 GHz luminosity density is in units of $10^{24}$ erg s$^{-1}$ Hz$^{-1}$ and the SFR is in units of $M_\odot$ year$^{-1}$.  We scale this relation to the luminosity density at 9 GHz assuming a spectral index of $\alpha=-0.5$ \citep{chomiukwilcots2009} and plot it with our measurements of $L_{\rm 9 GHz}$ (\S\ref{sec:radiosources}) vs.\ galaxy SFR (\S\ref{sec:HII}) in the middle panel of Figure \ref{fig:sne}.  Nearly all of our detected radio sources lie above the relation (including scatter), indicating that individual SNRs/SNe are not a viable explanation for the compact radio sources detected in our sample of dwarf galaxies.  

\begin{figure*}[t!]
\begin{center}
\hspace{-0.25cm}
\includegraphics[width=6.8in]{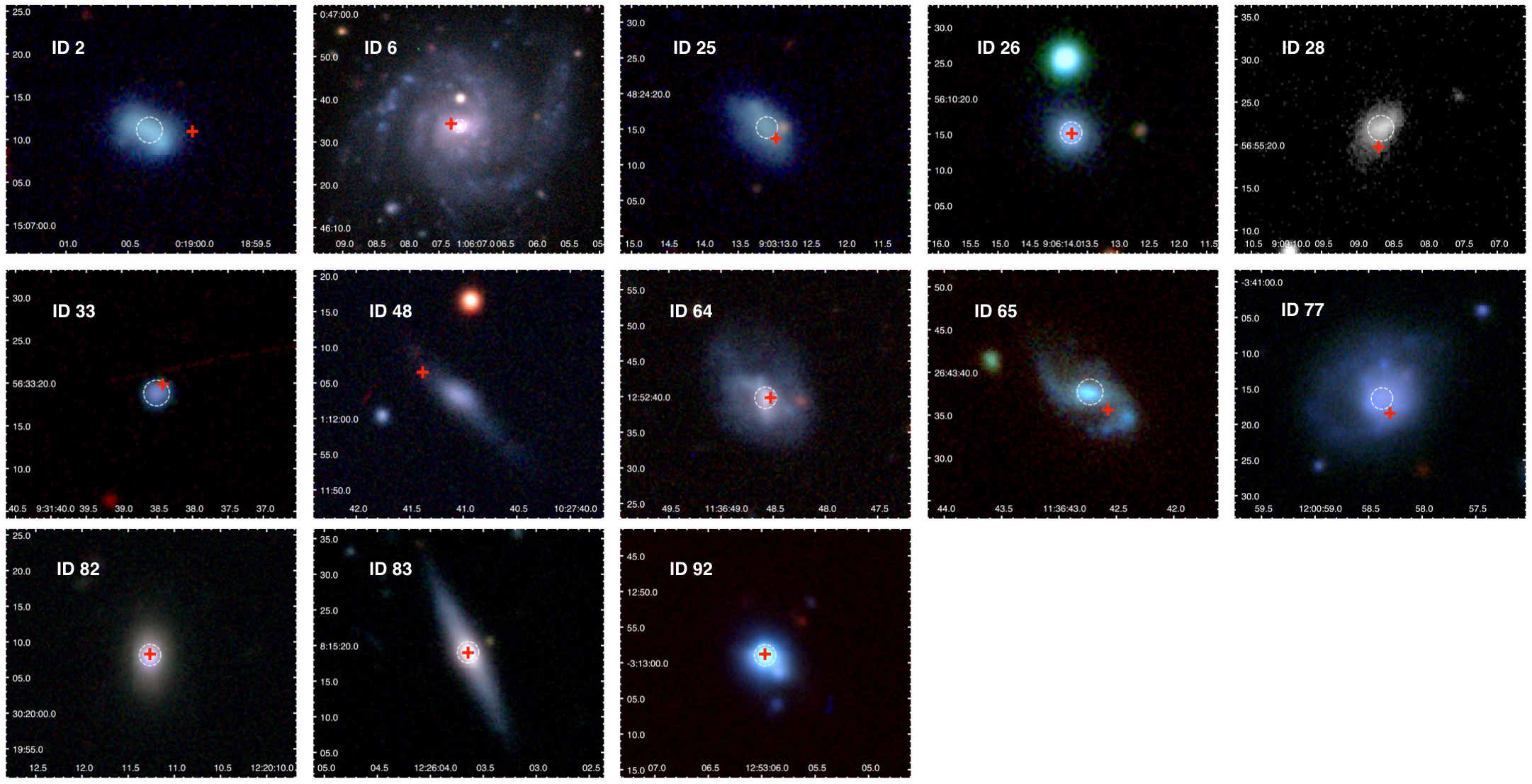}
\end{center}
\caption{DECaLS $grz$-band images of dwarf galaxies (in Sample A) with the strongest evidence for accreting massive BHs (see \S \ref{sec:agns}).  ID 28 only has a $z$-band image, which is shown here in grayscale.  Red crosses indicate the positions of the compact radio sources and the white dashed circles indicate the positions of the SDSS spectroscopic fibers (with diameters of $3\arcsec$).  ID 48 does not have an SDSS spectrum.  Absolute astrometry for the optical and radio images is accurate to $\lesssim 0\farcs1$ {and uncertainties in the radio source centroids are also $\lesssim 0\farcs1$.}
\label{fig:agns}}
\end{figure*}

\subsection{Populations of SNRs/SNe?}\label{sec:snpop}

We also consider multiple SNRs/SNe as a possible origin for the compact radio emission, making use of the radio SNR luminosity function derived  from a sample of 19 nearby galaxies spanning the SMC to Arp 220 \citep{chomiukwilcots2009}.  We adopt the luminosity function that is proportional to the host galaxy SFR, although we note that a population of SNRs/SNe producing the detected radio emission would need to be confined to roughly the size of VLA beam as the majority of our detections are point-like at a resolution of $\sim 0\farcs25$ (corresponding to a physical scale of $\sim 85$ pc at the median distance of our sample).  The luminosity function is given by    

\begin{equation} 
n(L) = \frac{dN}{dL} = 92 \times {\rm SFR} \times L^{-2.07},
\end{equation} 

\noindent
 where $n(L)$ is the number of SNRs with 1.4 GHz luminosity density $L$, and the SFRs of our sample galaxies are determined in Section \ref{sec:HII}.  The total expected luminosity from all SNRs in a given galaxy is then
 
\begin{equation} 
L_{\rm total} = \int n(L) ~L ~dL,
\end{equation}  

\noindent
where we take the integral from 0.1 to $10^4$ to span the full range of individual SNRs/SNe presented in \citet{chomiukwilcots2009}.  For each galaxy in our sample, we convert $L_{\rm total}$ (in units of $10^{24}$ erg s$^{-1}$ Hz$^{-1}$) to a luminosity at 9 GHz (in W Hz$^{-1}$) assuming a spectral index of $\alpha = -0.5$ \citep{chomiukwilcots2009}.  A comparison between the expected galaxy-wide luminosity of all SNRs and our detected compact radio sources is shown in the right panel of Figure \ref{fig:sne}.  Thirteen radio sources in Sample A have luminosities {more than a factor of 3 above} the expected cumulative contribution from all SNRs/SNe in their respective host galaxies, indicating an alternative origin for the radio emission is required.     

\begin{figure*}[!t]
\begin{center}
\hspace{-0.4cm}
\includegraphics[width=7.1in]{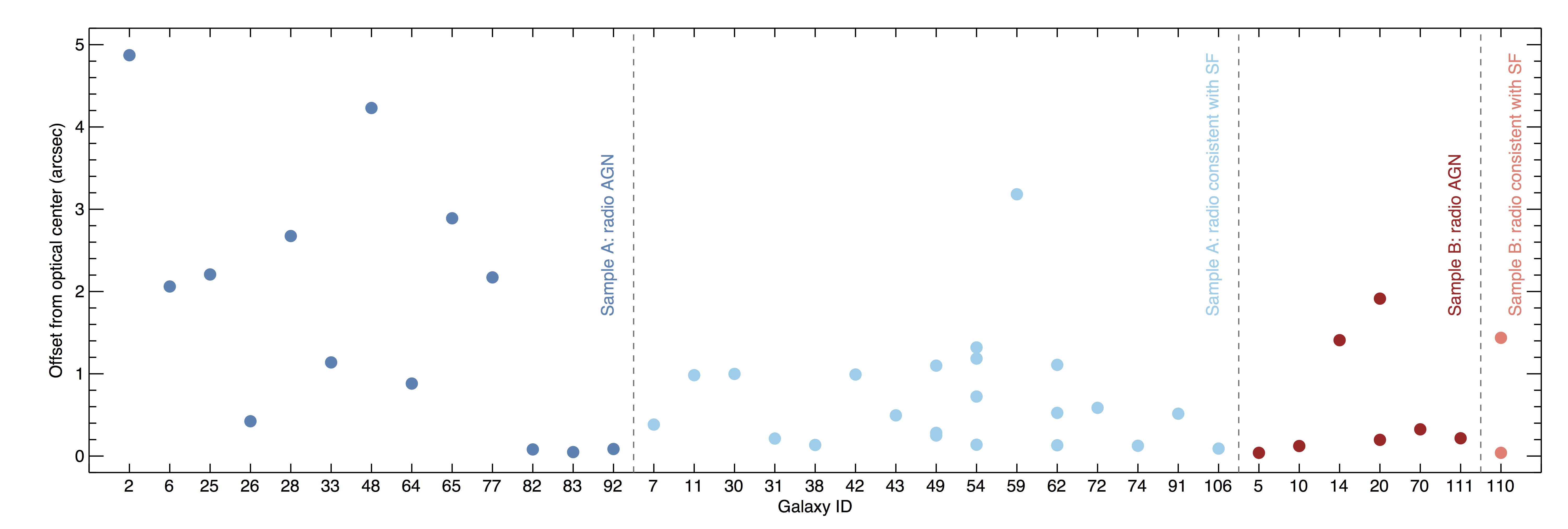}
\end{center}
\vspace{-.35cm}
\caption{Offset of radio sources relative to the optical centers of the target galaxies.  The SDSS, as well as the VLA observations, have absolute astrometry accurate to $\lesssim 0\farcs1$.  
\label{fig:offset_all}}
\end{figure*}

\begin{figure}[!h]
\begin{center}
\includegraphics[width=2.75in]{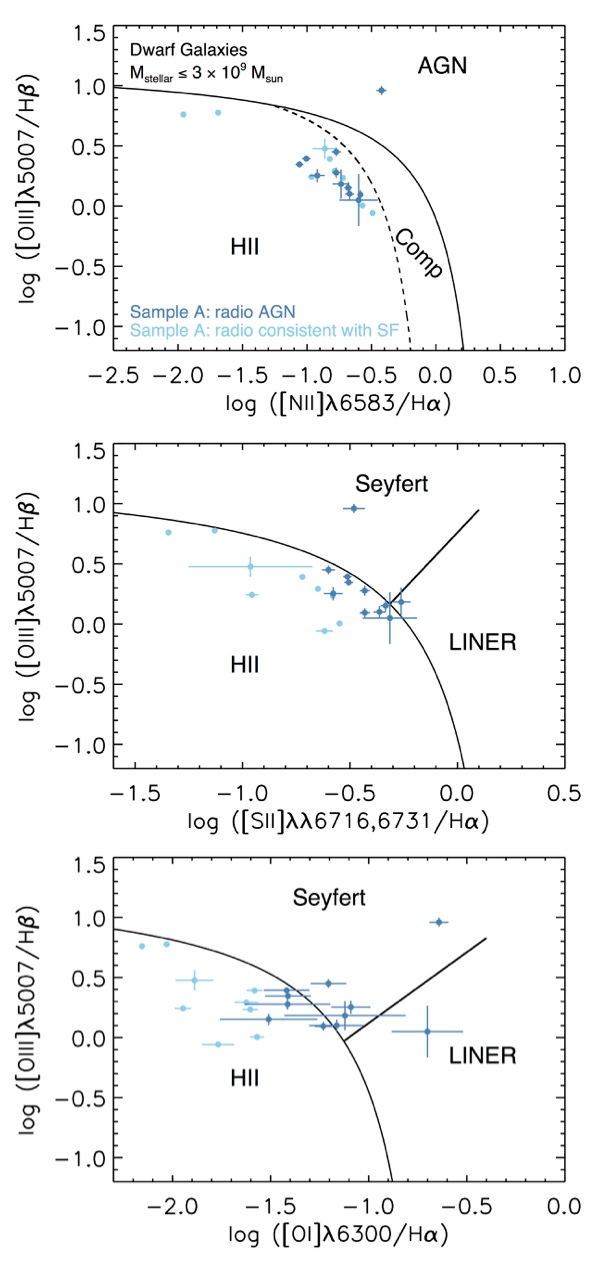}
\end{center}
\vspace{-0.25cm}
\caption{Optical emission line diagnostic diagrams for the galaxies in Sample A with SDSS spectra and emission line measurements from \citet{reinesvolonteri2015}.  The galaxies in Sample B do not have SDSS spectra.  The demarcations come from the classification scheme outlined in \citet{kewleyetal2006}.
\label{fig:bptplots}}
\end{figure}

\subsection{AGNs}\label{sec:agns}

Based on the analysis above, we have identified 13 compact radio sources in Sample A that are almost certainly AGNs. {Figure \ref{fig:agns} shows optical images of the host galaxies from the Dark Energy Camera Legacy Survey (DECaLS).}  The compact radio sources towards these dwarf galaxies are inconsistent with thermal \HII\ regions, individual SNRs/SNe, or populations of SNRs/SNe.  One of these objects, ID 26, was previously identified as a broad-line AGN in \citet{reinesetal2013} (designated RGG 9) with $M_{\rm BH} \sim 2.5 \times 10^5 M_\odot$.  
Additional compact radio sources detected in the other dwarf galaxies in Sample A may also be AGNs, however we cannot reliably rule out alternative explanations.  We discuss the objects in Sample B below in Section \ref{sec:sampB}.

{We investigate where the host galaxies in Sample A fall on optical diagnostic diagrams for those with SDSS spectra and narrow emission line measurements by \citet{reinesvolonteri2015}.  None of the galaxies in Sample B have SDSS spectra.}     
The narrow line diagnostic diagrams are shown in Figure \ref{fig:bptplots}.  ID 26 (RGG 9) is the only dwarf galaxy with emission line ratios clearly dominated by an AGN in all three diagrams.  The other sources fall in the star-forming region of the BPT diagram \citep{baldwinetal1981}, which takes \OIII/H$\beta$ versus \NII/H$\alpha$.  
However, 5 objects (IDs 25, 26, 64, 65, 83) fall in the Seyfert region of the \OIII/H$\beta$ versus \OI/H$\alpha$ diagram and 1 other object falls in the LINER region (ID 6).  These seemingly contradictory classifications do not preclude the existence of AGNs since \OI/H$\alpha$ is sensitive to the hardness of the radiation field, which moves AGN host galaxies to the right, while \NII/H$\alpha$ is primarily sensitive to metallicity and low mass galaxies are generally metal poor, which moves them to the left \citep[e.g., ][]{grovesetal2006,kewleyetal2006}.  We also note that the radio sources do not always fall within the SDSS spectroscopic fiber (see Figure \ref{fig:agns}) so these optical classifications should be taken with caution. 

Many of the radio-selected AGNs are not located in well-defined galaxy nuclei (Figures \ref{fig:agns} and \ref{fig:offset_all}), in contrast to the optically-selected AGNs in dwarf galaxies found by \citet{reinesetal2013}.  Roughly half of the galaxies have radio sources with offsets $\gtrsim 2\arcsec$ {relative to the optical centers provided in the NSA.  The absolute astrometry errors on both the optical and radio positions are $\lesssim 0\farcs1$}.  Some of the host dwarf galaxies do not even possess obvious photometric centers, having irregular morphologies and/or showing signs of interactions/mergers.  

We therefore carefully consider whether the 13 AGNs in Sample A reside in the target dwarf galaxies or are spurious background sources (also see \S\ref{sec:background} and \S\ref{sec:radiosources}).  Up to $\sim7$ background interlopers could be present, however there is compelling evidence that the radio AGNs in Sample A are indeed associated with the target galaxies.  First, unlike the background sources highlighted in Figure \ref{fig:sdss_backim}, the off-nuclear radio sources in Sample A lack detectable optical counterparts.  This argues against background interlopers, as they would have to be very obscured (particularly at DECaLS depths) or at high redshift.

\begin{figure*}[!h]
\begin{center}$
\begin{array}{cc}
\vspace{-.5cm}
\includegraphics[width=3in]{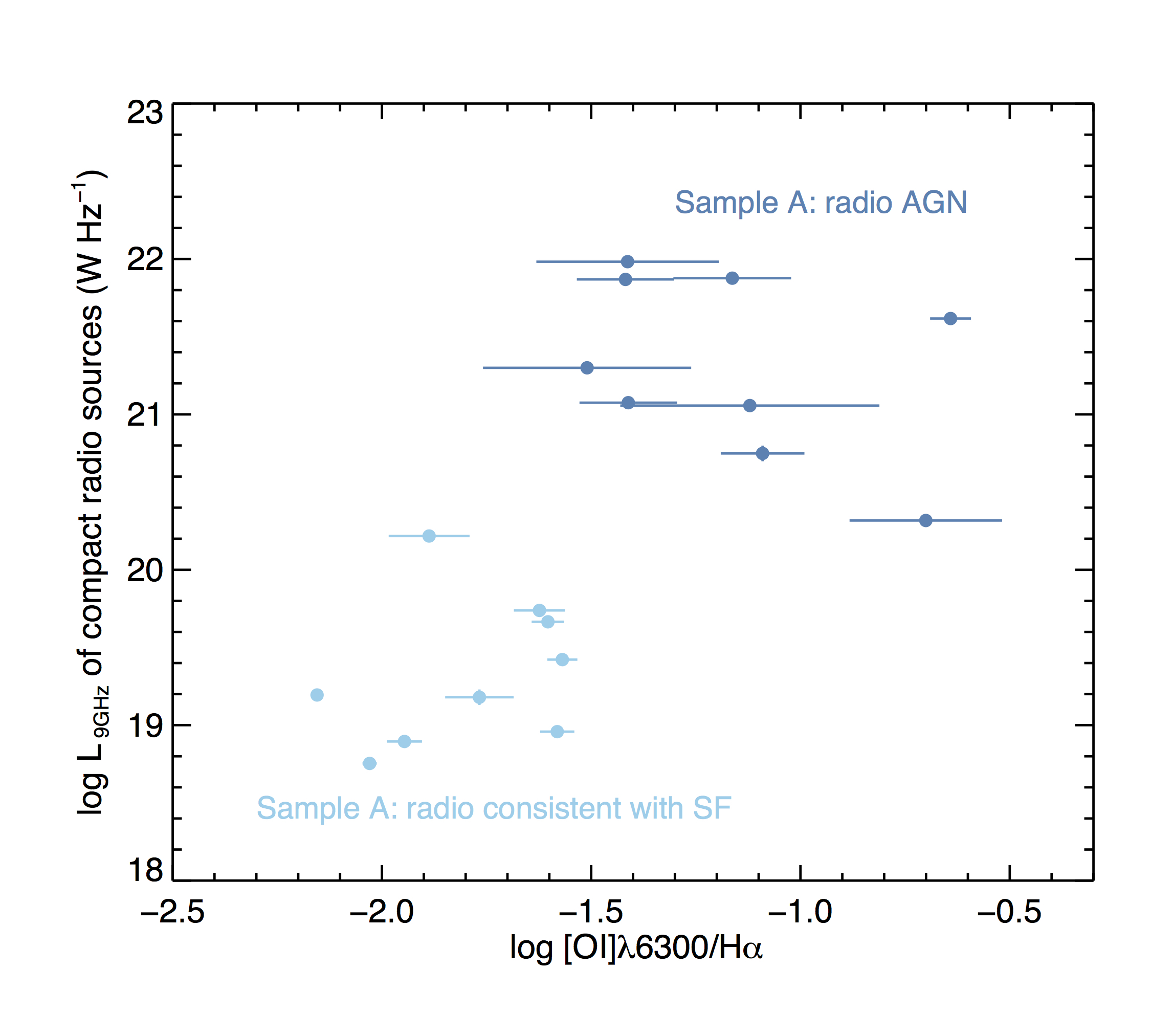} &
\includegraphics[width=3in]{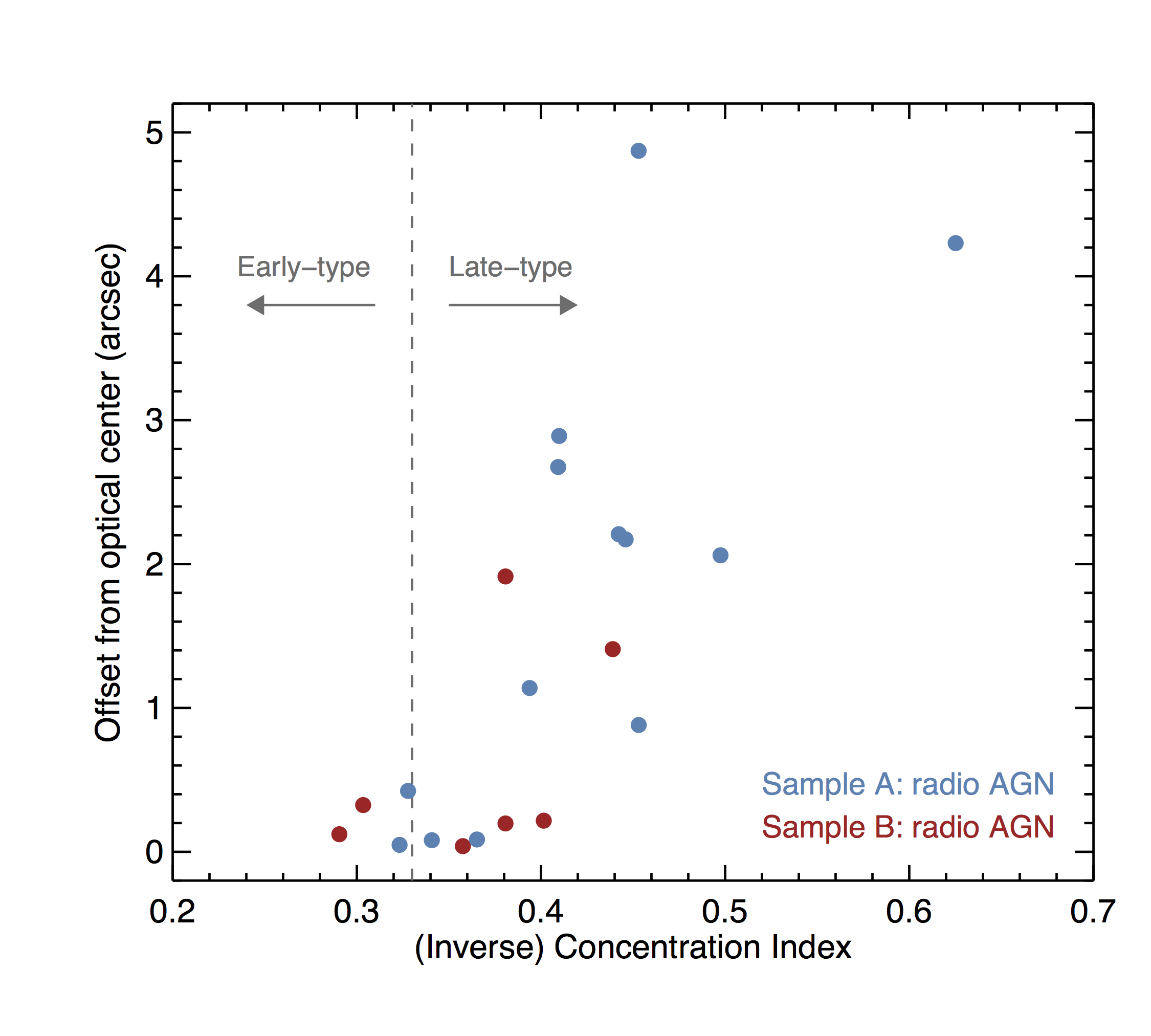}
\end{array}$
\end{center}
\caption{Left: Luminosity density at 9 GHz versus \OI/H$\alpha$ for dwarf galaxies in Sample A with SDSS spectra and emission line measurements from \citet{reinesvolonteri2015}.  The radio AGN have systematically higher \OI/H$\alpha$ ratios, suggesting the radio sources are associated with the dwarf galaxies producing the optical line emission, rather than background interlopers.
Right:  Offset of radio source from optical center of galaxy versus (inverse) concentration index defined as $C = r_{50}/r_{90}$ \citep{shimasakuetal2001}.  The more centrally concentrated galaxies have radio AGNs near their centers, while more extended/disturbed galaxies tend to host offset radio AGNs. 
\label{fig:o1ha}}
\end{figure*}

Additionally, Figure \ref{fig:o1ha} (left panel) shows that the dwarf galaxies in Sample A with radio-selected AGNs have systematically higher ratios of \OI/H$\alpha$ relative to the other dwarfs in Sample A that have less luminous radio sources consistent with star formation.  Combined with the aforementioned Seyfert/LINER classifications, this suggests that the radio sources are likely associated with the galaxies that are producing the optical line emission rather than background interlopers.  

We also see a trend such that the more extended/disturbed (later-type) galaxies tend to have more offset radio sources, while the centrally concentrated (earlier-type) galaxies tend to have nuclear radio sources (see Figure \ref{fig:agns}).  This is quantified in the right panel of Figure \ref{fig:o1ha}, which shows the radio-optical positional offset versus (inverse) concentration index defined as $C = r_{50}/r_{90}$, where $r_{50}$ and $r_{90}$ are the half-light and 90\% light Petrosian radii, respectively. In other words, the radio sources tend to fall in the nucleus of nucleated dwarf galaxies, whereas the radio sources are more offset for those galaxies without prominent nuclei.  This is consistent with simulations that predict a population of ``wandering" BHs in dwarf galaxies, primarily due to interactions/mergers \citep{bellovaryetal2019}.  

\begin{figure*}[!t]
\begin{center}
\hspace{-.2cm}
\includegraphics[width=7in]{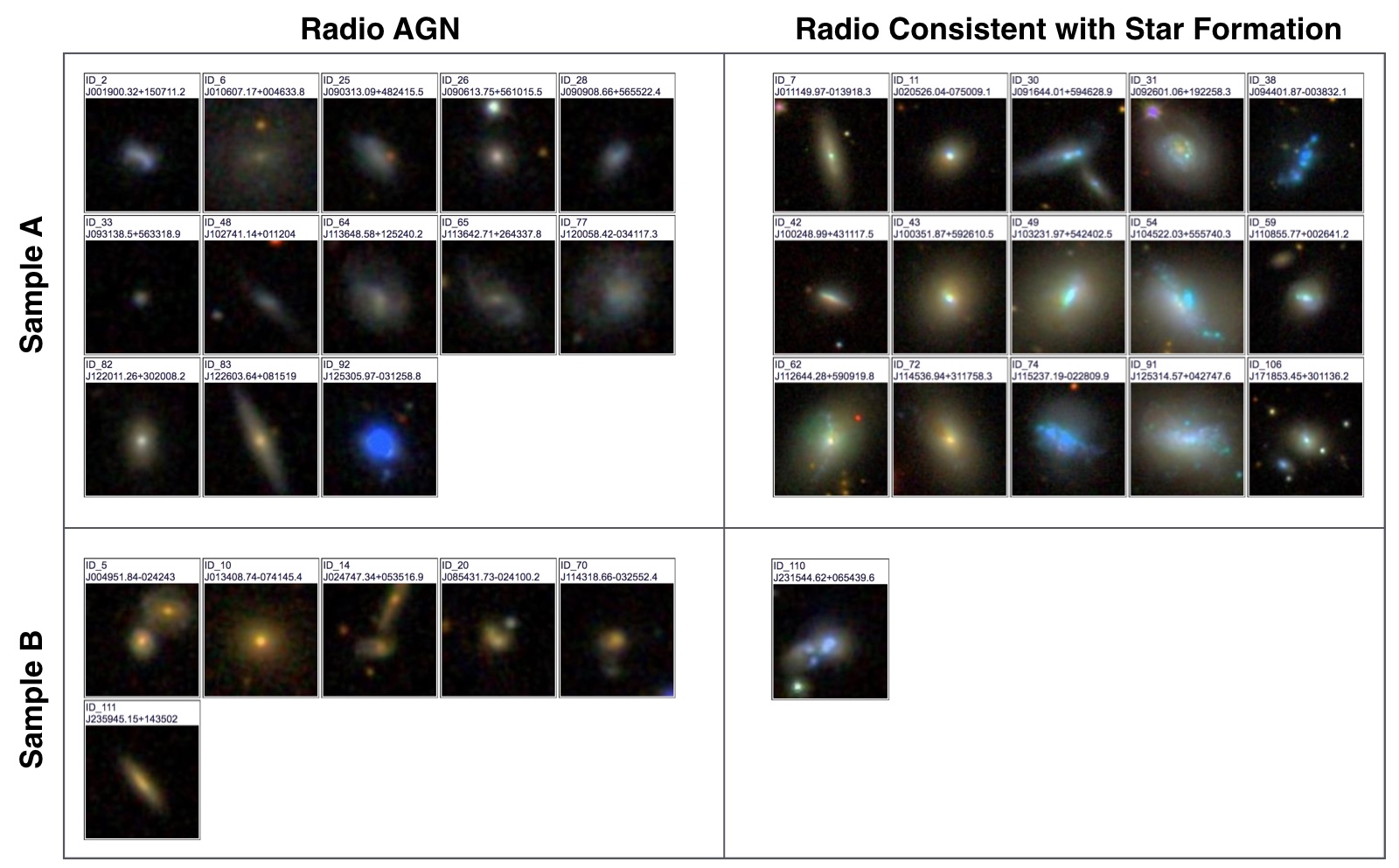}
\end{center}
\caption{SDSS images of the galaxies with compact radio sources detected in our VLA observations.  All of the compact radio sources are within 5\arcsec\ from the optical center of the galaxy.  See Table \ref{tab:sample} for galaxy properties.  Images are 25\arcsec\ on a side except the ones with radio sources consistent with star formation, which are 50\arcsec\ on a side.
\label{fig:sdss_ims}}
\end{figure*}

\subsubsection{Additional Objects in Sample B Including a Candidate Dual AGN}\label{sec:sampB}

We detect radio AGNs in 6 additional galaxies in Sample B (Figure \ref{fig:sdss_ims}).  SDSS spectra are not available for the galaxies in Sample B and since we do not have reliable spectroscopic redshifts, the distances and masses are uncertain for these galaxies.  Moreover, we cannot investigate where they fall on the optical diagnostic diagrams.  

Photometric redshifts are available from the SDSS DR14 Sky Server and fall in the range $0.08 \lesssim z_{\rm phot} \lesssim 0.16$ (with a median value of 0.1).  These values are significantly larger than the redshifts in the NSA ($0.005 \lesssim z_{\rm phot} \lesssim 0.016$, median value of 0.01). Along with the redder colors relative to the dwarf galaxies in Sample A (see Figure \ref{fig:sdss_ims}), this suggests these galaxies are much more massive than the values provided in the NSA ($M_{\rm stellar} \lesssim 2 \times 10^9 M_\odot$).

\begin{figure}[!t]
\begin{center}
\vspace{.5cm}
\hspace{-0.4cm}
\includegraphics[width=3in]{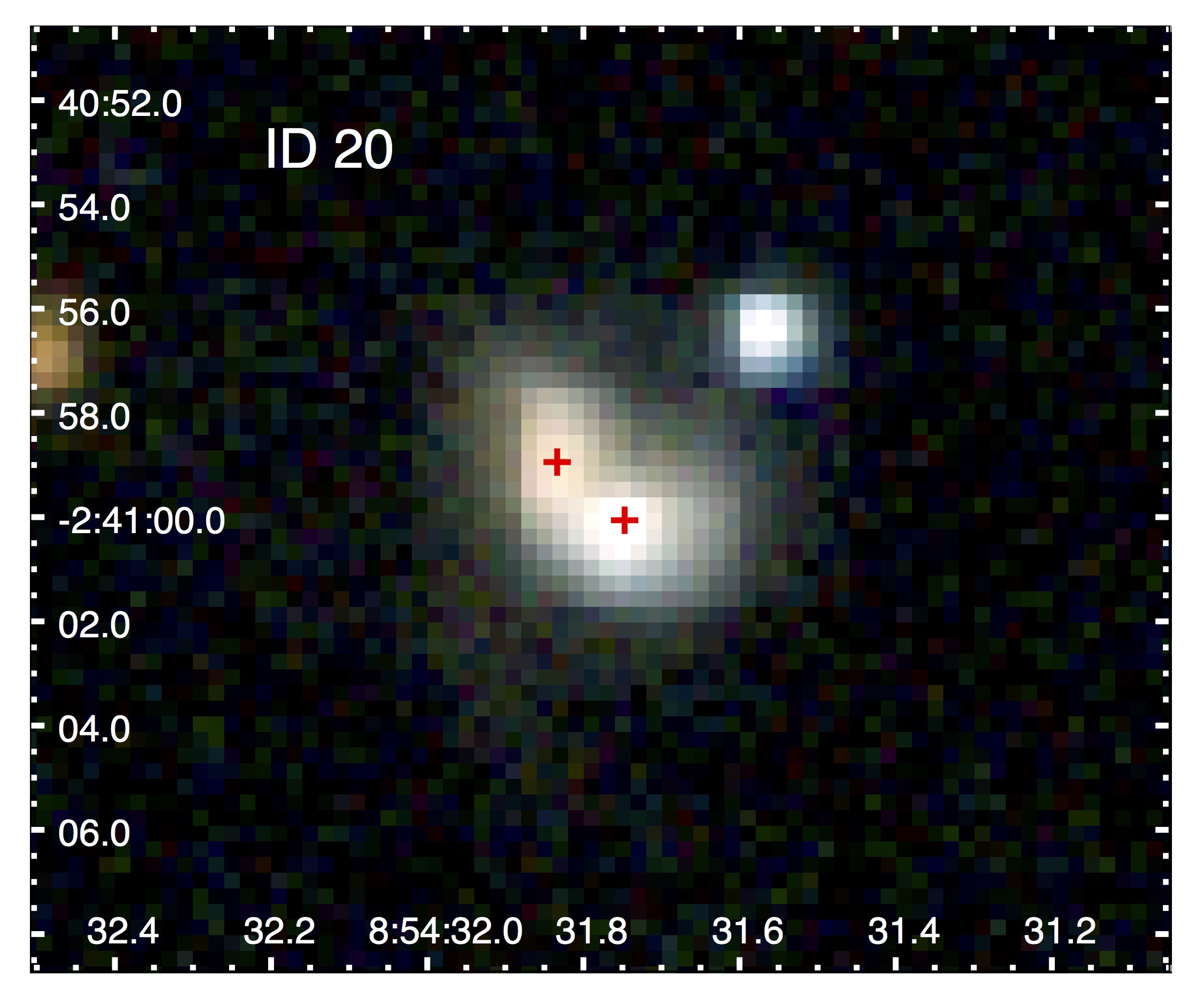}
\end{center}
\caption{DECaLS image of the dual radio AGN candidate ID 20.  The red crosses indicate the positions of the VLA sources.
\label{fig:id20}}
\end{figure}

We detect two radio AGNs towards ID 20.  Figure \ref{fig:id20} shows a DECaLS image of the system, which appears to consist of two galaxies.  The radio sources we detect are approximately 1\farcs7 apart and coincide with the optical galaxies.  Under the assumption that galaxies are at the same redshift ($z_{\rm phot} \sim 0.16$), this corresponds to a physical projected separation of $\sim 5.5$ kpc.  Therefore, ID 20 is a strong dual AGN candidate \citep[e.g.,][]{comerfordetal2015}, one of only a handful of those detected in the radio \citep[e.g.,][]{fuetal2015,mullersanchezetal2015}.

\section{Properties of the Dwarf Galaxies Hosting Radio-Selected AGNs}

\begin{figure}[!t]
\begin{center}
\hspace{-0.4cm}
\includegraphics[width=3.5in]{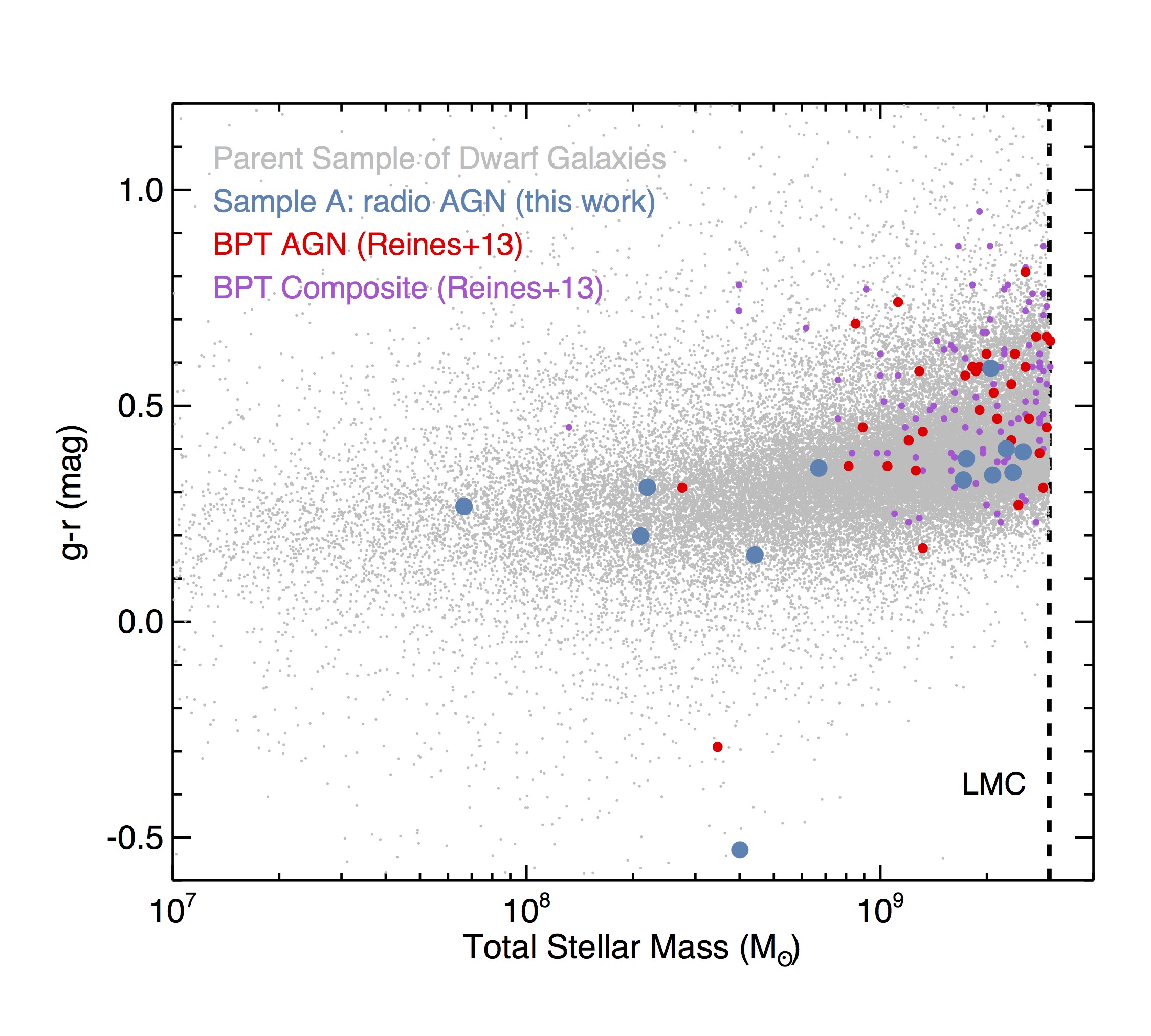}
\end{center}
\vspace{-.5cm}
\caption{Plot of $g-r$ color versus total stellar mass for dwarf galaxies with $M_\star \leq 3 \times 10^{9}~M_\odot$ in the NSA.  The radio-detected AGNs in this work are shown along with the optically-selected AGNs and composites from \citet{reinesetal2013}.
\label{fig:colormass}}
\end{figure}

We obtain galaxy half-light radii, total stellar masses and colors and from the NSA (Table \ref{tab:sample}). The galaxies in Sample A with radio-selected AGNs are physically small.  The half-light radii span a range of $\sim 0.5-10$ kpc, with the majority (10/13) having $r_{\rm 50} \lesssim 3$ kpc.  Total stellar masses are in the range $M_\star \sim 10^{7.8} - 10^{9.4}~ M_\odot$ with a median value of $M_\star \sim 10^{9.2}~M_\odot$.  Three of the galaxies have stellar masses below $M_\star \lesssim 3 \times 10^{8}~M_\odot$, which is the mass of the lowest-mass BPT AGN from \citet{reinesetal2013} and equivalent to the stellar mass of the SMC \citep{stanimirovicetal2004}. The radio-detected dwarf galaxies follow the overall color distribution of the general population of dwarf galaxies, whereas the optically-selected sample of AGNs in dwarfs from Reines et al.\ (2013) tend to have redder host galaxies (see Figure \ref{fig:colormass}).  This illustrates the advantage of using high-resolution radio observations to overcome some of the bias against finding AGNs in bluer star-forming dwarf galaxies that optical selection suffers.  

\begin{figure}[!t]
\begin{center}
\hspace{-0.4cm}
\includegraphics[width=3.5in]{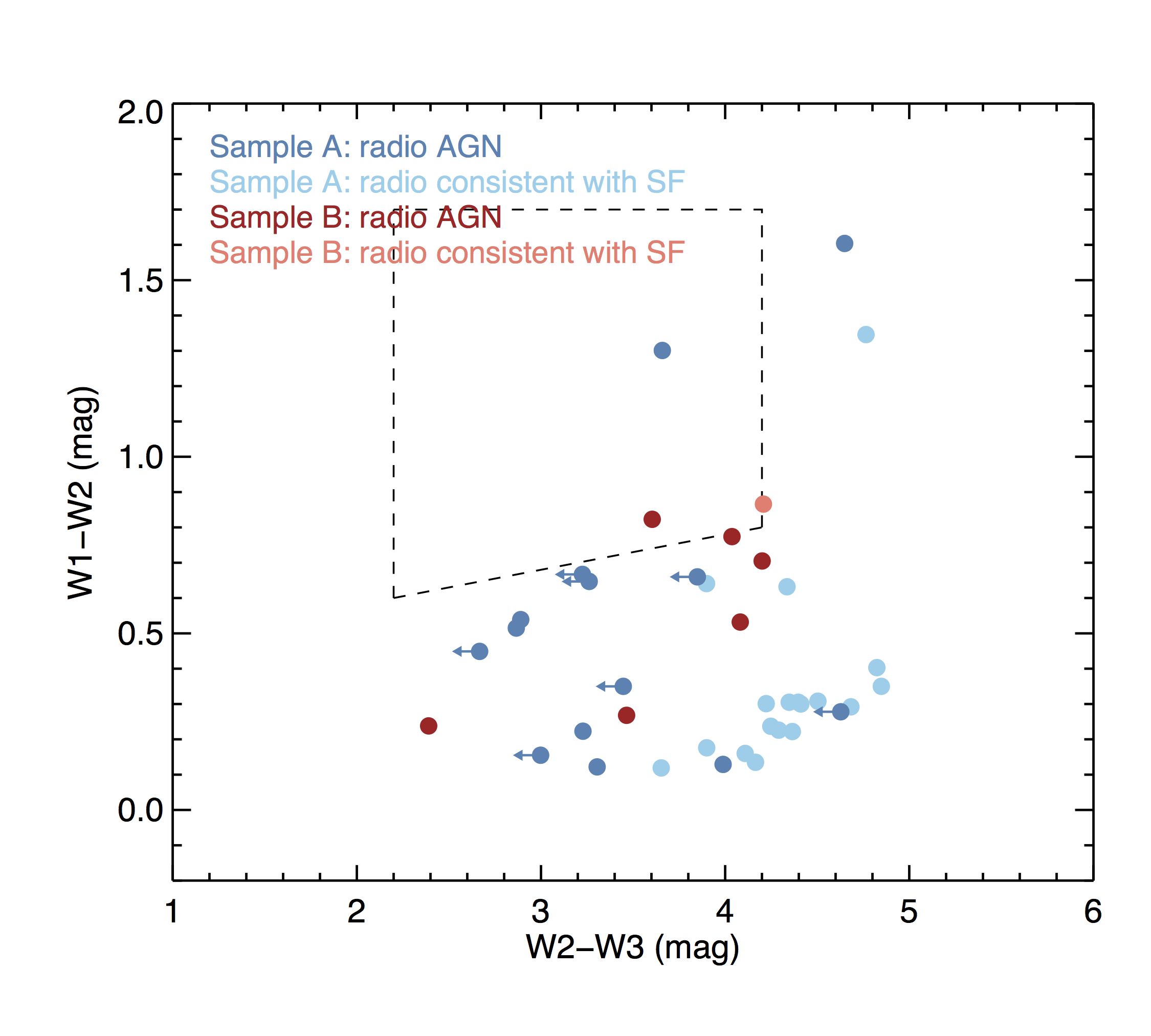}
\end{center}
\vspace{-.5cm}
\caption{{\it WISE} color-color diagram for the galaxies in our sample.  The \citet{jarrettetal2011} AGN selection box is shown with dashed lines.
\label{fig:wise}}
\end{figure}

Mid-infrared colors of the galaxies in our sample are shown in Figure \ref{fig:wise}.  We use the {\it WISE} magnitudes from the AllWISE Source Catalog to see if any of the sources show signs of AGN heated dust.  ID 26 (RGG 9) is the only dwarf galaxy that falls in the \citet{jarrettetal2011} AGN selection box.  The majority of the objects in Sample A are dominated by infrared emission from the host galaxies.  A couple of additional objects would be selected by the \citet{sternetal2012} AGN selection criterion, $W1-W2 \geq 0.8$.  While this single color selection has proved useful for luminous sources, \citet{hainlineetal2016} caution that dwarf starburst galaxies can severely contaminate samples at low stellar masses.  

\section{Conclusions and Discussion}

Here we present the first large-scale radio search for massive BHs in nearby dwarf galaxies ($M_\star \lesssim 3 \times 10^9 M_\odot$; also see \citealt{mezcuaetal2019} for a search at intermediate-redshifts using VLA-COSMOS).  We observe 111 dwarf galaxies detected in the FIRST survey using new VLA continuum observations at X-band with $\sim$10 times the sensitivity and a beam solid angle that is smaller by a factor of $\sim$400 compared to FIRST.  Thirteen dwarf galaxies have compact radio sources that are almost certainly AGNs, as they are inconsistent with alternative explanations for the radio emission (e.g., thermal \HII\ regions, individual SNRs/SNe, populations of SNRs/SNe). We also find evidence suggesting the radio AGNs are indeed associated with the dwarf galaxies, rather than background interlopers (e.g., Figure \ref{fig:o1ha}).

Overall the host galaxies are bluer than dwarfs with optically-selected AGNs \citep{reinesetal2013}, they have narrow emission line ratios falling in the star-forming region of the \OIII/H$\beta$ versus \NII/H$\alpha$ BPT diagram, and they have mid-IR colors dominated by stellar emission.  With one exception (ID 26, RGG 9), the AGNs we have found in dwarf galaxies using high-resolution radio observations are effectively hidden at optical/IR wavelengths.

The BH masses in these systems are largely unknown.  However, if we assume the BH mass to total stellar mass relation derived from optically-selected broad-line AGNs \citep{reinesvolonteri2015}, we would expect BH masses in the range of $M_{\rm BH} \sim 10^{4.1} - 10^{5.8}~ M_\odot$ with a median value of $M_{\rm BH} \sim 10^{5.6}~M_\odot$.  We caution that there are large uncertainties associated with these mass estimates. The scatter in the \citet{reinesvolonteri2015} AGN relation is 0.55 dex and it is not clear that this relation is applicable to our host galaxies, particularly for any galaxies undergoing interactions/mergers.

Many of the radio AGNs we identify are not located in well-defined nuclei and some are significantly offset from the center of their host galaxy (Figure \ref{fig:agns}).  This is distinct from the known dwarf galaxies hosting optically-selected AGNs \citep[e.g.,][]{reinesetal2013,schutteetal2019}.  While surprising from an observational standpoint, recent simulations predict that roughly half of all massive BHs in dwarf galaxies are expected to be wandering around in the outskirts of their host galaxies \citep{bellovaryetal2019}.  Off-nuclear BHs may have been tidally stripped during mergers \citep[e.g.,][]{governatoetal1994,bellovaryetal2010,tremmeletal2015}, or they could be recoiling merged BHs \citep[e.g.,][]{merrittetal2004,volonteriperna2005}.  The coalescence of two BHs will produce gravitational waves that carry momentum, causing the merged black hole to undergo a velocity kick and spatial displacement. Gravitational waves from massive BH mergers in dwarf galaxies are expected to be detectable with future observations with {\it LISA}.  {Once a massive black hole has left the center of a dwarf galaxy, it is unlikely to return. This is in contrast to more massive galaxies where dynamical friction is more efficient and can bring the black hole back to the nucleus.}

We are currently pursuing a broad range of follow-up observations of our sample.  For example, spatially-resolved spectroscopy covering the positions of the radio sources will help confirm that they are not background objects, and VLBI observations may resolve the radio emission or confine it to a more compact core.  Additionally, high-resolution X-ray observations will help solidify the AGN nature of the radio sources and can be used with the fundamental plane of BH activity to estimate BH masses and Eddington ratios.     

Finally, this work highlights the potential of deep, high-resolution radio observations for searching for (wandering) massive BHs in dwarf galaxies. Ultimately we need a larger, unbiased radio survey (e.g., with a next generation VLA; \citealt{plotkinreines2018})
to better constrain the occupation fraction of massive BHs in dwarf galaxies and thereby determine the mechanism that seeded the first high-redshift BHs.  

\acknowledgements

AER is extremely grateful to Amy Kimball, J\"urgen Ott, and Drew Medlin at NRAO in Socorro for helping with the VLA data reduction.  The authors also thank Andy Goulding for helpful discussions about galaxy ID 47 and the referee for a constructive report. The National Radio Astronomy Observatory is a facility of the National Science Foundation operated under cooperative agreement by Associated Universities, Inc.  This publication makes use of data products from the Wide-field Infrared Survey Explorer, which is a joint project of the University of California, Los Angeles, and the Jet Propulsion Laboratory/California Institute of Technology, funded by NASA. This work has also used observations made with the NASA Galaxy Evolution Explorer. GALEX is operated for NASA by the California Institute of Technology under NASA contract NAS5--98034. This study has made use of the NASA/IPAC Extragalactic Database (NED), which is operated by the Jet Propulsion Laboratory, California Institute of Technology, under contract with NASA.  We are grateful to Michael Blanton and all who helped create the NASA-Sloan Atlas. Funding for the NASA-Sloan Atlas has been provided by the NASA Astrophysics Data Analysis Program (08-ADP08-0072) and the NSF (AST-1211644).

Funding for the Sloan Digital Sky Survey IV has been provided by the Alfred P. Sloan Foundation, the U.S. Department of Energy Office of Science, and the Participating Institutions. SDSS-IV acknowledges
support and resources from the Center for High-Performance Computing at
the University of Utah. The SDSS web site is www.sdss.org.
SDSS-IV is managed by the Astrophysical Research Consortium for the 
Participating Institutions of the SDSS Collaboration including the 
Brazilian Participation Group, the Carnegie Institution for Science, 
Carnegie Mellon University, the Chilean Participation Group, the French Participation Group, Harvard-Smithsonian Center for Astrophysics, 
Instituto de Astrof\'isica de Canarias, The Johns Hopkins University, Kavli Institute for the Physics and Mathematics of the Universe (IPMU) / 
University of Tokyo, the Korean Participation Group, Lawrence Berkeley National Laboratory, 
Leibniz Institut f\"ur Astrophysik Potsdam (AIP),  
Max-Planck-Institut f\"ur Astronomie (MPIA Heidelberg), 
Max-Planck-Institut f\"ur Astrophysik (MPA Garching), 
Max-Planck-Institut f\"ur Extraterrestrische Physik (MPE), 
National Astronomical Observatories of China, New Mexico State University, 
New York University, University of Notre Dame, 
Observat\'ario Nacional / MCTI, The Ohio State University, 
Pennsylvania State University, Shanghai Astronomical Observatory, 
United Kingdom Participation Group,
Universidad Nacional Aut\'onoma de M\'exico, University of Arizona, 
University of Colorado Boulder, University of Oxford, University of Portsmouth, 
University of Utah, University of Virginia, University of Washington, University of Wisconsin, 
Vanderbilt University, and Yale University.

The Legacy Surveys consist of three individual and complementary projects: the Dark Energy Camera Legacy Survey (DECaLS; NOAO Proposal ID \# 2014B-0404; PIs: David Schlegel and Arjun Dey), the Beijing-Arizona Sky Survey (BASS; NOAO Proposal ID \# 2015A-0801; PIs: Zhou Xu and Xiaohui Fan), and the Mayall z-band Legacy Survey (MzLS; NOAO Proposal ID \# 2016A-0453; PI: Arjun Dey). DECaLS, BASS and MzLS together include data obtained, respectively, at the Blanco telescope, Cerro Tololo Inter-American Observatory, National Optical Astronomy Observatory (NOAO); the Bok telescope, Steward Observatory, University of Arizona; and the Mayall telescope, Kitt Peak National Observatory, NOAO. The Legacy Surveys project is honored to be permitted to conduct astronomical research on Iolkam Du'ag (Kitt Peak), a mountain with particular significance to the Tohono O'odham Nation.
NOAO is operated by the Association of Universities for Research in Astronomy (AURA) under a cooperative agreement with the National Science Foundation.
This project used data obtained with the Dark Energy Camera (DECam), which was constructed by the Dark Energy Survey (DES) collaboration. Funding for the DES Projects has been provided by the U.S. Department of Energy, the U.S. National Science Foundation, the Ministry of Science and Education of Spain, the Science and Technology Facilities Council of the United Kingdom, the Higher Education Funding Council for England, the National Center for Supercomputing Applications at the University of Illinois at Urbana-Champaign, the Kavli Institute of Cosmological Physics at the University of Chicago, Center for Cosmology and Astro-Particle Physics at the Ohio State University, the Mitchell Institute for Fundamental Physics and Astronomy at Texas A\&M University, Financiadora de Estudos e Projetos, Fundacao Carlos Chagas Filho de Amparo, Financiadora de Estudos e Projetos, Fundacao Carlos Chagas Filho de Amparo a Pesquisa do Estado do Rio de Janeiro, Conselho Nacional de Desenvolvimento Cientifico e Tecnologico and the Ministerio da Ciencia, Tecnologia e Inovacao, the Deutsche Forschungsgemeinschaft and the Collaborating Institutions in the Dark Energy Survey. The Collaborating Institutions are Argonne National Laboratory, the University of California at Santa Cruz, the University of Cambridge, Centro de Investigaciones Energeticas, Medioambientales y Tecnologicas-Madrid, the University of Chicago, University College London, the DES-Brazil Consortium, the University of Edinburgh, the Eidgenossische Technische Hochschule (ETH) Zurich, Fermi National Accelerator Laboratory, the University of Illinois at Urbana-Champaign, the Institut de Ciencies de l'Espai (IEEC/CSIC), the Institut de Fisica d'Altes Energies, Lawrence Berkeley National Laboratory, the Ludwig-Maximilians Universitat Munchen and the associated Excellence Cluster Universe, the University of Michigan, the National Optical Astronomy Observatory, the University of Nottingham, the Ohio State University, the University of Pennsylvania, the University of Portsmouth, SLAC National Accelerator Laboratory, Stanford University, the University of Sussex, and Texas A\&M University.
BASS is a key project of the Telescope Access Program (TAP), which has been funded by the National Astronomical Observatories of China, the Chinese Academy of Sciences (the Strategic Priority Research Program ``The Emergence of Cosmological Structures" Grant \# XDB09000000), and the Special Fund for Astronomy from the Ministry of Finance. The BASS is also supported by the External Cooperation Program of Chinese Academy of Sciences (Grant \# 114A11KYSB20160057), and Chinese National Natural Science Foundation (Grant \# 11433005).
The Legacy Survey team makes use of data products from the Near-Earth Object Wide-field Infrared Survey Explorer (NEOWISE), which is a project of the Jet Propulsion Laboratory/California Institute of Technology. NEOWISE is funded by the National Aeronautics and Space Administration.
The Legacy Surveys imaging of the DESI footprint is supported by the Director, Office of Science, Office of High Energy Physics of the U.S. Department of Energy under Contract No. DE-AC02-05CH1123, by the National Energy Research Scientific Computing Center, a DOE Office of Science User Facility under the same contract; and by the U.S. National Science Foundation, Division of Astronomical Sciences under Contract No. AST-0950945 to NOAO.

\appendix

\subsection{ID 47 - A Serendipitous Discovery of an Optical AGN in a Dwarf Galaxy}\label{sec:id47} 

We detect an off-nuclear radio source towards the dwarf galaxy J1022$-$0055 (ID 47) with a luminosity much too high to be consistent with stellar processes.  The radio emission consists of a bright point source plus a fainter elongated jet-like structure (Figure \ref{fig:vla_ims}), features commonly associated with accreting massive BHs.  The radio source is located 4\farcs9 away from the center of the galaxy in the north-east direction, and has a point-like optical counterpart that can be seen in ground-based imaging (Figure \ref{fig:sdss_backim}). 

\begin{figure*}[!h]
\begin{center}
\hspace{-0.3cm}
\includegraphics[width=7in]{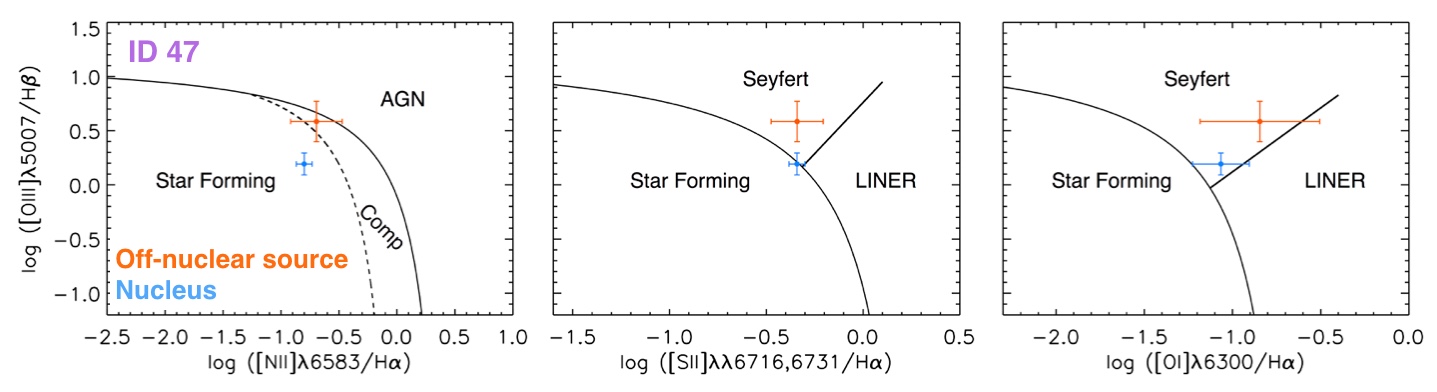}
\end{center}
\caption{Optical emission-line diagnostic diagrams showing the nucleus and off-nuclear source in galaxy ID 47. 
\vspace{.2cm}
\label{fig:id47_bpt}}
\end{figure*}

SDSS spectra were obtained for both the off-nuclear source and the galaxy center.  Based on the narrow emission lines, the offset source has a redshift of $z = 0.04828 \pm 0.00003$ and the galaxy center has a redshift of $z = 0.04853 \pm 0.00001$, indicating a radial velocity offset of $\Delta v = 75$ km s$^{-1}$ with a statistical error of 10 km s$^{-1}$.  In other words, the gas producing the narrow emission lines at the position of the offset source is associated with the dwarf galaxy.  Moreover, our emission line measurements from the optical spectrum suggest the presence of an accreting massive black hole.  The source falls in the Seyfert regions of the \OIII/H$\beta$ versus \SII/H$\alpha$ and \OIII/H$\beta$ versus \OI/H$\alpha$ diagnostic diagrams, as well as the AGN/star-forming composite region of the \OIII/H$\beta$ versus \NII/H$\alpha$ BPT diagram (Figure \ref{fig:id47_bpt}).  

\begin{figure}[!h]
\begin{center}
\hspace{-0.4cm}
\includegraphics[width=4in]{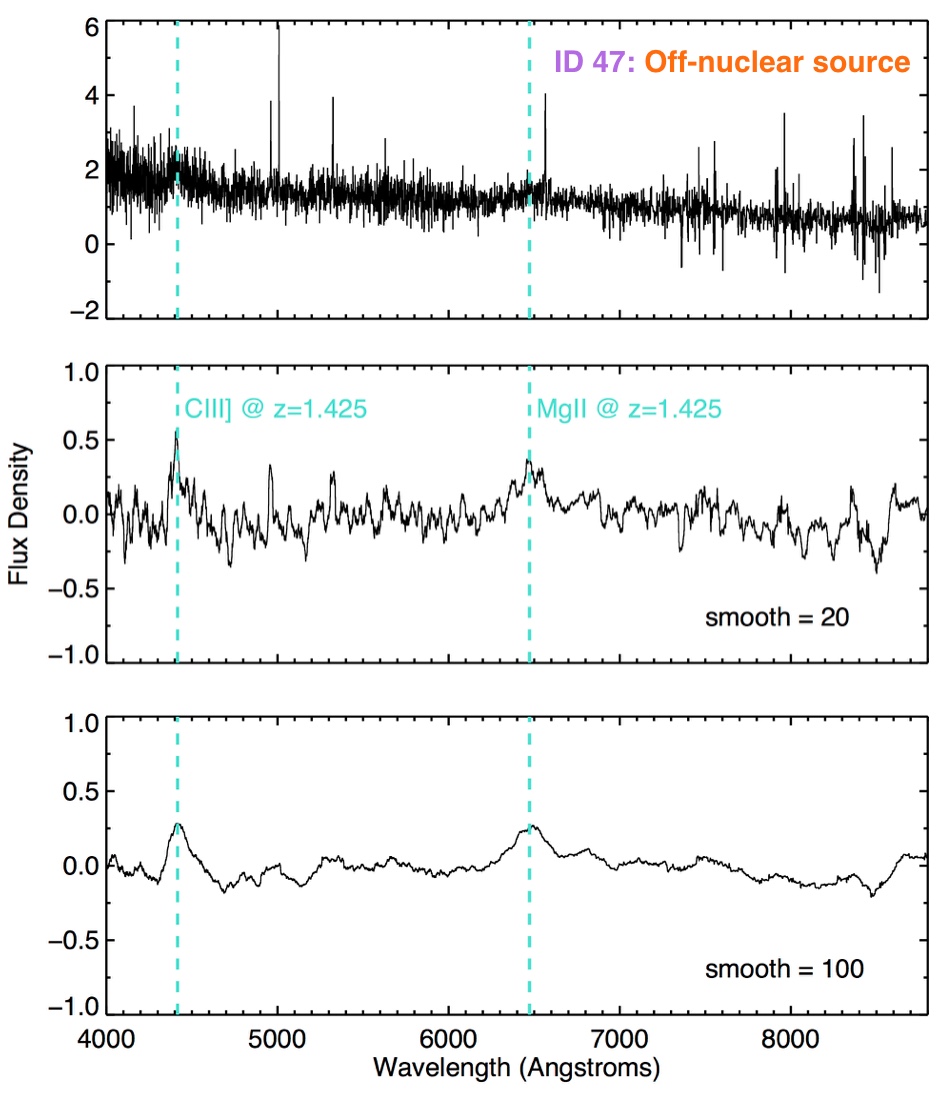}
\end{center}
\caption{Top: SDSS spectrum of the off-nuclear radio source in ID 47 (Figure \ref{fig:sdss_backim}) at the rest wavelength of the galaxy (z=0.48).  Middle: The same spectrum after subtracting the continuum and strong emission lines, smoothed with a boxcar average of 20 pixels.  Bottom: Same as middle panel, but smoothed with a boxcar average of 100 pixels.  The turquoise dashed lines indicate redshifted broad UV lines commonly associated with quasars. 
\label{fig:id47_quasar}}
\end{figure}

While at first glance these findings suggest the presence of an offset massive BH in the dwarf galaxy, a closer inspection of the optical spectrum reveals broad, low-level emission lines that are consistent with CIII]$\lambda 1909$ and MgII$\lambda 2798$ at a redshift of $z=1.425$ (see Figure \ref{fig:id47_quasar}).  This suggests the offset source is more likely a background quasar, which can account for the broad lines, the radio source and the optical counterpart.  The optical counterpart of the off-nuclear radio source is a variable point source in ground-based imaging. SDSS images of J1022$-$0055 were taken on March 22, 1999 and the DR15 catalog $g$- and $r$-band magnitudes of the off-nuclear source are $g = 21.83 \pm 0.07$ mag and $r = 21.22 \pm 0.06$ mag\footnote{http://skyserver.sdss.org/dr15/en/tools/quicklook/summary.aspx?ra=10:22:27.86\&dec=$-$00:55:30.36}. Images of J1022$-$0055 were also obtained as part of the Dark Energy Camera Legacy Survey (DECaLS) beginning in March 2015 with the last $g$-band image taken in February 2018 and the last $r$-band image taken in March 2017. The $g$- and $r$- band magnitudes of the off-nuclear source in the DECaLS DR7 catalog (from the stacked images) are $g = 22.98 \pm 0.02$ mag and $r = 22.02 \pm 0.01$ mag. Therefore, the off-nuclear source dimmed by $\sim$1 mag between the SDSS and DECaLS imaging observations that were taken $\sim$ 16 years apart.

The narrow emission lines in the spectrum of the offset source can be attributed to gas in the dwarf galaxy that is ionized by an AGN, although the BH need not be at the location of the offset source.  The nuclear spectrum of the galaxy shows hints of an AGN in the diagnostic diagrams shown in Figure \ref{fig:id47_bpt}.  Therefore, it is plausible that there is indeed an AGN at the center of the dwarf galaxy, where the line emission due to accretion in the nuclear region is diluted by star formation but is more readily seen in the outskirts of the galaxy \citep[e.g.,][]{wylezaleketal2018}. Nevertheless, it is somewhat surprising to find serendipitous optical evidence for an AGN in a dwarf galaxy that was selected because of a chance alignment with an unrelated radio source, potentially signaling an appreciable BH occupation fraction for galaxies of comparable mass ($M_\star \sim 3 \times 10^9~M_\odot \sim M_{\star, \rm LMC}$).  

\vspace{.4cm}

\clearpage

\begin{deluxetable*}{ccrrrcccrrccrr}
\tabletypesize{\footnotesize}
\tablecaption{Observed Dwarf Galaxy Sample with FIRST Detections}
\tablewidth{0pt}
\tablehead{
\colhead{} & \colhead{} & \multicolumn{9}{c}{NSA (optical)} & \colhead{}  & \multicolumn{2}{c}{FIRST (radio)} \\
\cline{3-11} \cline{13-14} \\
\colhead{ID} & \colhead{Name} & \colhead{NSAID}  & \colhead{R.A.} & \colhead{Decl.} & 
\colhead{$z$}  & \colhead{log M$_\star$} & \colhead{$M_g$} & \colhead{$g-r$} & \colhead{$r_{50}$}  & \colhead{S{\'e}rsic $n$} &
\colhead{} & \colhead{$S_{\rm 1.4GHz}$} & \colhead{Offset}   \\
\colhead{(1)} & \colhead{(2)} & \colhead{(3)} & \colhead{(4)} & \colhead{(5)} & \colhead{(6)} & \colhead{(7)} & \colhead{(8)} &
\colhead{(9)} & \colhead{(10)} & \colhead{(11)} & \colhead{} & \colhead{(12)} & \colhead{(13)} }
\startdata
1 & J0018$-$0903 & 21913 & 4.53817 & $-9.05993$ & 0.0233 & 9.05 & $-18.42$ & $0.45$ & 0.77 & 5.9 & & 0.71 & 0.7 \\
2 & J0019+1507 & 26027 & 4.75134 & $15.11978$ & 0.0376 & 8.65 & $-18.42$ & $0.15$ & 1.51 & 0.8 & & 2.58 & 4.3 \\
3 & J0032$-$0656 & 127158 & 8.20313 & $-6.94158$ & 0.0162 & 9.42 & $-18.65$ & $0.39$ & 1.09 & 1.0 & & 2.49 & 2.4 \\
4 & J0048+0037 & 62105 & 12.12363 & $0.61966$ & 0.0383 & 8.40 & $-17.67$ & $-0.03$ & 0.95 & 0.5 & & 2.86 & 3.9 \\
5 & J0049$-$0242 & 127796 & 12.46603 & $-2.71196$ & 0.0131 & 9.01 & $-16.75$ & $0.68$ & 0.47 & 5.0 & & 0.96 & 1.3 \\
6 & J0106+0046 & 23750 & 16.52989 & $0.77606$ & 0.0171 & 9.40 & $-18.42$ & $0.39$ & 5.01 & 1.8 & & 1.47 & 2.2 \\
7 & J0111$-$0139 & 128822 & 17.95825 & $-1.65511$ & 0.0158 & 9.46 & $-19.12$ & $0.52$ & 1.29 & 3.7 & & 0.73 & 0.9 \\
8 & J0115$-$1002 & 22511 & 18.75235 & $-10.03804$ & 0.0232 & 8.98 & $-19.08$ & $0.14$ & 1.45 & 1.0 & & 1.58 & 1.3 \\
9 & J0118+1212 & 129053 & 19.50442 & $12.20686$ & 0.0193 & 9.35 & $-19.13$ & $0.13$ & 0.66 & 3.6 & & 1.03 & 1.2 \\
10 & J0134$-$0741 & 130010 & 23.53643 & $-7.69597$ & 0.0156 & 9.28 & $-16.87$ & $1.07$ & 0.94 & 6.0 & & 20.62 & 0.1
\enddata
\tablecomments{Table \ref{tab:sample} is published in its entirety in the electronic edition of {\it The Astrophysical Journal}.
A portion is shown here for guidance regarding its form and content.  
Column 1: galaxy identification number assigned in this paper.
Column 2: galaxy name.
Column 3: NSA identification number.
Column 4: right ascension in units of degrees. 
Column 5: declination in units of degrees.    
Column 6: redshift.
Column 7: log galaxy stellar mass in units of M$_\odot$.
Column 8: absolute $g$-band magnitude.
Column 9: $g-r$ color.
Column 10: Petrosian 50\% light radius in units of kpc. 
Column 11: S{\'e}rsic index, $n$. 
Values in columns 4-11 are from the NSA and assume $h=0.73$.
Magnitudes are $K$-corrected to rest-frame values using \texttt{kcorrect v4\_2} and corrected for foreground Galactic extinction.
Column 12: radio flux density in units of mJy at 1.4 GHz from the FIRST survey.
Column 13: offset from optical position of galaxy in units of arcseconds.}
\label{tab:sample}
\end{deluxetable*}

\begin{deluxetable*}{ccccccrcccrc}
\tabletypesize{\footnotesize}
\tablecaption{VLA Observations}
\tablewidth{0pt}
\tablehead{
\colhead{} & \colhead{} &\colhead{} & \colhead{} &\colhead{} & \multicolumn{3}{c}{9 GHz}  & \colhead{}  & \multicolumn{3}{c}{10.65 GHz} \\
\cline{6-8} \cline{10-12} \\
\colhead{ID} & \colhead{Name}  & \colhead{Date}  & \colhead{Flux} & \colhead{Phase} & \colhead{Synthesized} & \colhead{P.A.} & \colhead{RMS} & \colhead{} & \colhead{Synthesized} & \colhead{P.A.} & \colhead{RMS} \\
\colhead{} & \colhead{} & \colhead{Observed} & \colhead{Calibrator}  & \colhead{Calibrator} & \colhead{Beam}  & \colhead{}  & \colhead{Noise}  &  \colhead{} & \colhead{Beam}  & \colhead{}  & \colhead{Noise} }
\startdata
1 & J0018$-$0903 & 2014-02-21 & 3C48 & J0006$-$0623 & $0.28 \times 0.17$ & $38$ & 15 &  & $0.24 \times 0.15$ & $41$ & 17 \\
2 & J0019+1507 & 2014-04-29 & 3C48 & J0010+1724 & $0.21 \times 0.17$ & $-84$ & 14 &  & $0.17 \times 0.15$ & $-82$ & 15 \\
3 & J0032$-$0656 & 2014-02-21 & 3C48 & J0006$-$0623 & $0.26 \times 0.18$ & $41$ & 16 &  & $0.22 \times 0.15$ & $44$ & 17 \\
4 & J0048+0037 & 2014-03-09 & 3C48 & J0115$-$0127 & $0.24 \times 0.18$ & $65$ & 24 &  & $0.21 \times 0.16$ & $69$ & 28 \\
5 & J0049$-$0242 & 2014-03-09 & 3C48 & J0115$-$0127 & $0.26 \times 0.18$ & $59$ & 15 &  & $0.22 \times 0.16$ & $59$ & 29  \\
6 & J0106+0046 & 2014-03-09 & 3C48 & J0115$-$0127 & $0.24 \times 0.18$ & $73$ & 14 &  & $0.20 \times 0.16$ & $74$ & 17 \\
7 & J0111$-$0139 & 2014-03-09 & 3C48 & J0115$-$0127 & $0.25 \times 0.19$ & $61$ & 14 &  & $0.21 \times 0.16$ & $62$ & 17 \\
8 & J0115$-$1002 & 2014-02-21 & 3C48 & J0116$-$1136 & $0.25 \times 0.17$ & $27$ & 16 &  & $0.21 \times 0.15$ & $28$ & 17 \\
9 & J0118+1212 & 2014-02-21 & 3C48 & J0121+0422 & $0.19 \times 0.17$ & $-86$ & 15 &  & $0.16 \times 0.15$ & $-83$ & 16 \\
10 & J0134$-$0741 & 2014-03-09 & 3C48 & J0116$-$1136 & $0.27 \times 0.18$ & $30$ & 20 &  & $0.23 \times 0.16$ & $31$ & 20
\enddata
\tablecomments{Table \ref{tab:vla} is published in its entirety in the electronic edition of {\it The Astrophysical Journal}.
A portion is shown here for guidance regarding its form and content. 
All observations were taken at X band while the VLA was in the A-configuration.  Synthesized beam sizes are in units of arcsec $\times$ arcsec.
Beam position angles are in units of degrees.  RMS noise has units of $\mu$Jy beam$^{-1}$.} 
\label{tab:vla}
\end{deluxetable*}

\begin{deluxetable*}{lcrrrrcrcl}
\tabletypesize{\footnotesize}
\tablecaption{Compact Radio Sources Detected Towards Target Galaxies}
\tablewidth{0pt}
\tablehead{
\colhead{ID} & \colhead{Name} & \colhead{R.A.} & \colhead{Decl.} & \colhead{Offset} & \colhead{$S_{\rm 9GHz}$} & 
\colhead{log $L_{\rm 9GHz}$} & \colhead{$\alpha$}  & \colhead{Point source} & \colhead{Note}  \\
\colhead{(1)} & \colhead{(2)} & \colhead{(3)} & \colhead{(4)} & \colhead{(5)} & \colhead{(6)} & \colhead{(7)} & \colhead{(8)} & \colhead{(9)} 
& \colhead{(10)}  }
\startdata
2 & J0019+1507 & 4.74994 & $15.11973$ & 4.9 & 2586(27) & 21.9 & $-0.2(0.1)$ & True & Sample A, AGN \\
5 & J0049$-$0242 & 12.46602 & $-2.71195$ & 0.0 & 372(34) & 20.1 & $-1.9(0.9)$ & False & Sample B, AGN \\
6 & J0106+0046 & 16.53045 & $0.77620$ & 2.1 & 352(24) & 20.3 & $-0.5(0.6)$ & True & Sample A, AGN \\
7 & J0111$-$0139 & 17.95820 & $-1.65520$ & 0.4 & 172(20) & 19.9 & $-1.2(1.1)$ & False & Sample A, SF \\
10 & J0134$-$0741 & 23.53642 & $-7.69593$ & 0.1 & 21831(200) & 22.0 & $0.1(0.1)$ & True & Sample B, AGN \\
11 & J0205$-$0750 & 31.35838 & $-7.83610$ & 1.0 & 126(37) & 19.6 & $-2.7(1.9)$ & True & Sample A, SF \\
14 & J0247+0535 & 41.94719 & $5.58766$ & 1.4 & 1148(31) & 19.7 & $-1.0(0.3)$ & True & Sample B, AGN \\
20a & J0854$-$0240 & 133.63228 & $-2.68335$ & 0.2 & 379(30) & 20.3 & $-0.8(0.9)$ & False & Sample B, AGN\\
20b & \nodata & 133.63264 & $-2.68304$ & 1.9 & 199(28) & 20.0 & $-2.0(1.5)$ & True & Sample B, AGN \\
25 & J0903+4824 & 135.80403 & $48.40381$ & 2.2 & 375(43) & 20.7 & $-1.3(1.2)$ & False & Sample A, AGN \\
26 & J0906+5610 & 136.55737 & $56.17087$ & 0.4 & 929(26) & 21.6 & $-1.0(0.3)$ & True & Sample A, AGN \\
28 & J0909+5655 & 137.28621 & $56.92215$ & 2.7 & 595(22) & 21.1 & $-1.1(0.4)$ & True & Sample A, AGN \\
30 & J0916+5946 & 139.18317 & $59.77447$ & 1.0 & 65(21) & 19.4 & \nodata & True & Sample A, SF \\
31 & J0926+1923 & 141.50447 & $19.38290$ & 0.2 & 63(19) & 19.0 & \nodata & True & Sample A, SF \\
33 & J0931+5633 & 142.91008 & $56.55552$ & 1.1 & 1951(27) & 22.0 & $-1.0(0.1)$ & True & Sample A, AGN \\
38 & J0944$-$0038 & 146.00783 & $-0.64223$ & 0.1 & 259(10) & 19.2 & $-1.3(0.4)$ & False & Sample A, SF \\
42 & J1002+4311 & 150.70386 & $43.18802$ & 1.0 & 66(23) & 19.7 & \nodata & True & Sample A, SF \\
43 & J1003+5926 & 150.96643 & $59.43628$ & 0.5 & 70(30) & 19.2 & $-0.8(3.2)$ & True & Sample A, SF \\
46 & J1018$-$0216 & 154.54269 & $-2.27437$ & 4.0 & 5444(33) & 22.4 & $-0.1(0.1)$ & True & Background  \\
47 & J1022$-$0055 & 155.61610 & $-0.92510$ & 4.9 & 1559(39) & 21.9 & $-1.4(0.2)$ & False & Background \\
48 & J1027+0112 & 156.92241 & $1.20179$ & 4.2 & 2587(33) & 21.4 & $-0.3(0.1)$ & True & Sample A, AGN \\
49a & J1032+5424 & 158.13327 & $54.40040$ & 1.1 & 108(22) & 18.9 & $-3.1(2.0)$ & True & Sample A, SF \\
49b & \nodata & 158.13329 & $54.40063$ & 0.3 & 102(43) & 18.9 & $0.0(3.8)$ & False & Sample A, SF \\
49c & \nodata & 158.13327 & $54.40077$ & 0.2 & 100(15) & 18.9 & $-0.4(0.9)$ & False & Sample A, SF \\
50 & J1034+0449 & 158.72845 & $4.82818$ & 2.8 & 375(26) & 20.8 & $-1.4(0.8)$ & True & Background \\
54a & J1045+5557 & 161.34189 & $55.96118$ & 0.1 & 458(64) & 19.2 & $-0.8(1.1)$ & False & Sample A, SF \\
54b & \nodata & 161.34129 & $55.96098$ & 1.3 & 340(30) & 19.1 & $-1.7(0.6)$ & False & Sample A, SF \\
54c & \nodata & 161.34145 & $55.96095$ & 1.2 & 103(30) & 18.5 & $-1.6(1.9)$ & False & Sample A, SF \\
54d & \nodata & 161.34158 & $55.96105$ & 0.7 & 99(40) & 18.5 & $-0.9(4.2)$ & False & Sample A, SF \\
59 & J1108+0026 & 167.23329 & $0.44485$ & 3.2 & 141(26) & 19.7 & $-1.0(1.6)$ & True & Sample A, SF \\
62a & J1126+5909 & 171.68445 & $59.15566$ & 0.5 & 173(31) & 19.0 & $-1.5(1.5)$ & False & Sample A, SF \\
62b & \nodata & 171.68434 & $59.15581$ & 1.1 & 140(26) & 18.9 & $-0.2(2.2)$ & False  & Sample A, SF \\
62c & \nodata & 171.68451 & $59.15549$ & 0.1 & 75(28) & 18.6 & $-2.0(2.8)$ & True & Sample A, SF \\
64 & J1136+1252 & 174.20219 & $12.87775$ & 0.9 & 3218(27) & 21.9 & $0.1(0.1)$ & True & Sample A, AGN  \\
65 & J1136+2643 & 174.17740 & $26.72657$ & 2.9 & 517(24) & 21.1 & $-1.9(0.5)$ & False & Sample A, AGN \\
70 & J1143$-$0325 & 175.82785 & $-3.43129$ & 0.3 & 1009(46) & 19.8 & $-1.3(0.5)$ & False & Sample B, AGN \\
72 & J1145+3117 & 176.40404 & $31.29970$ & 0.6 & 301(41) & 19.4 & $-1.4(1.4)$ & False & Sample A, SF \\
74 & J1152$-$0228 & 178.15494 & $-2.46945$ & 0.1 & 180(33) & 18.8 & $-1.2(1.7)$ & True & Sample A, SF \\
77 & J1200$-$0341 & 180.24292 & $-3.68846$ & 2.2 & 1496(33) & 21.3 & $-0.9(0.2)$ & True & Sample A, AGN \\
82 & J1220+3020 & 185.04694 & $30.33564$ & 0.1 & 397(24) & 20.8 & $-0.7(0.6)$ & True & Sample A, AGN \\
83 & J1226+0815 & 186.51518 & $8.25528$ & 0.0 & 780(26) & 21.0 & $-0.6(0.3)$ & True & Sample A, AGN \\
91 & J1253+0427 & 193.31082 & $4.46313$ & 0.5 & 312(30) & 18.7 & $-0.7(0.9)$ & True & Sample A, SF \\
92 & J1253$-$0312 & 193.27487 & $-3.21632$ & 0.1 & 1160(76) & 21.1 & $-0.8(0.6)$ & False & Sample A, AGN \\
96 & J1301+2809 & 195.40129 & $28.15157$ & 4.3 & 783(27) & 20.8 & $-1.3(0.3)$ & True & Background \\
106 & J1718+3011 & 259.72269 & $30.19341$ & 0.1 & 381(51) & 20.2 & $-1.6(1.2)$ & False & Sample A, SF \\
110a & J2315+0654 & 348.93619 & $6.91072$ & 1.4 & 250(10) & 19.5 & $-0.5(0.8)$ & False & Sample B, SF \\
110b & \nodata & 348.93591 & $6.91103$ & 0.0 & 112(10) & 19.1 & $0.4(1.1)$ & False & Sample B, SF \\
111 & J2359+1435 & 359.93811 & $14.58394$ & 0.2 & 442(26) & 20.3 & $-0.7(0.6)$ & False & Sample B, AGN
\enddata
\tablecomments{Column 1: galaxy ID (see Table \ref{tab:sample}). 
Column 2: galaxy name.
Column 3: right ascension of radio source in units of degrees. 
Column 4: declination of radio source in units of degrees.  
Column 5: offset from optical center of the galaxy in units of arcseconds.
Column 6: flux density at 9 GHz in units of $\mu$Jy.  Uncertainties are given in parenthesis. This does not include the 5\% calibration error.
Column 7: log spectral luminosity at 9 GHz in units of W Hz$^{-1}$. 
Column 8: spectral index ($S_\nu \propto \nu^\alpha$) determined from flux densities at 9.00 GHz and 10.65 GHz,
if source is detected in the 10.65 GHz image.
Uncertainties are given in parenthesis.  
Column 9: designation as a point source or not determined from IMFIT.
Column 10: indication of whether the radio source is in Sample A, Sample B, or is a background source. ``AGN" indicates radio sources that are too luminous to be consistent with star formation.
``SF" indicates radio sources consistent with star formation.}
\label{tab:radio}
\end{deluxetable*}

\end{document}